\documentstyle[prd,aps]{revtex}
\begin{document}
 \input epsf
\draft
\renewcommand{\topfraction}{0.8}
\twocolumn[\hsize\textwidth\columnwidth\hsize\csname
@twocolumnfalse\endcsname
\preprint{UH-IfA-29, hep-ph/9705347, April 26, 1997}
\title {\bf Structure of Resonance in Preheating after Inflation   }
\author{Patrick B. Greene}
\address{ Department of Physics and Astronomy, University of Hawaii,
2505 Correa Rd., Honolulu HI 96822, USA}
\author{Lev Kofman}
\address{ Institute for Astronomy, University of Hawaii,
2680 Woodlawn Dr., Honolulu, HI 96822, USA }
\author{Andrei Linde}
\address{ Department of Physics, Stanford University, Stanford CA
94305-4060, USA}
\author {Alexei A. Starobinsky}
\address{ Landau Institute for Theoretical Physics,
Kosygina St. 2, Moscow 117334, Russia}
\date { May 18, 1997}
\maketitle
\begin{abstract}
We consider preheating in the theory
${1 \over 4} \lambda \phi^4  + {1 \over 2} g^2 \phi^2 \chi^2 $, where
the classical oscillating  inflaton field $\phi(t)$ decays into $\chi$-particles and $\phi$-particles.
 The parametric resonance
which leads to particle production  in this  conformally invariant theory is described by the
Lame equation. It significantly  differs  from the resonance in the theory with a quadratic potential.
 The structure of the resonance   depends in a rather nontrivial way on the
parameter $g^2/\lambda$. We construct the stability/instability
chart in this theory for   arbitrary  $g^2/\lambda$. We  give simple  analytic solutions describing the resonance in the limiting cases $g^2/\lambda\ll 1$ and $g^2/\lambda \gg 1$, and 
in   the theory with $g^2=  3\lambda$, and   with  $g^2 = \lambda$. From the point of view of parametric resonance for $\chi$, the theories with $g^2=  3\lambda$  and   with  $g^2 = \lambda$
  have the same structure, respectively, as the theory ${1 \over 4} \lambda \phi^4 $, and   the theory ${\lambda\over 4 N}(\phi^2_i)^2$ of an $N$-component scalar field $\phi_i$ in the limit $N \to \infty$.
We show that in some of the conformally invariant theories  such as the simplest 
model ${1 \over 4} \lambda \phi^4 $,  the resonance can
be terminated by the backreaction of produced particles long before  $\langle \chi^2 \rangle$ or
 $\langle   \phi^2 \rangle $ become of the order   $\phi^2$. We analyze the changes in the theory of reheating in this model which appear if the inflaton field has a small mass. 
 
\end{abstract}
\pacs{PACS: 98.80.Cq  \hskip 2.5 cm IfA-97-29~~~~~~SU-ITP-97-19 \hskip 2.5 cm 
hep-ph/9705347}
 \vskip2pc]

\section{Introduction}
 
The theory of reheating of the universe is one of the most important and least developed parts of inflationary cosmology. Recently it was found that in many realistic versions of chaotic inflation reheating begins with a stage of parametric resonance \cite{KLS}. At this stage the energy is rapidly transferred from the inflaton field to other scalar and vector fields interacting with it. This process occurs far away from thermal equilibrium, and therefore we called it {\it preheating}. The theory of preheating is rather complicated. In \cite{KLS} we gave only a brief summary of its basic features.  A detailed investigation of preheating in the simplest chaotic inflation model describing a massive inflaton field $\phi$ interacting with a massless scalar field $\chi$ was contained in our recent paper \cite{KLS97}. It was found, in particular, that the resonance in such theories can be efficient only if it is extremely broad. In such a situation preheating in an expanding universe looks like a stochastic process.  

In this paper we will concentrate on the theory of preheating in a class of conformally invariant theories such as ${\lambda\over 4}\phi^4 + {g^2\over 2}\phi^2\chi^2$.
Different aspects of preheating in such theories   have been
 studied in Refs. \cite{KLS,Shtanov,Kof96,KhTk961,Boyan95,Boyan96,Kaiser97,rt,Prokopec}.
A specific feature of these models is that by a conformal transformation one can  reduce the investigation of preheating in these theories in an expanding universe to a much simpler theory of preheating  in Minkowski space-time \cite{KLS}. As a result, the parametric resonance does not exhibit the stochasticity found in \cite{KLS97}. However, stochastic resonance may appear again at the late stages of preheating if the fields $\phi$ and $\chi$ have bare masses which break conformal invariance.

We will investigate preheating in the theories of the type of    ${\lambda\over 4}\phi^4 + {g^2\over 2}\phi^2\chi^2$ for various relations between the coupling constants $g^2$ and $\lambda$. During this investigation (see specifically Sec. V and XII),
 we will discuss how the results of the  previous papers on this subject are related to
the picture which emerges from the current study.
We will show that the development of the resonance  in the various conformally invariant
theories
can  be very different, depending on the particular values
of parameters and the structure
of the theory.
For example,  the model  ${\lambda\over 4}\phi^4 + {g^2\over 2}\phi^2\chi^2$  with $g^2 = \lambda$ or $g^2 = 3\lambda$  has only one instability band, but the structure of the bands  and the characteristic exponents  $\mu_k$  are completely different.  It is enough to change    the ratio   $g^2/\lambda$ only slightly, and the number of the instability bands immediately becomes infinitely large. For this reason, it is dangerous to extrapolate the results  obtained for a theory with one choice of parameters
to a  theory with another choice of parameters.
As we will see, not only is the structure of the resonances   different
in different models, but the self-consistent dynamical evolution with an account taken of the backreaction of produced particles 
can also be qualitatively different.

The main purpose of the present paper is to
study the structure of the parametric resonance in the conformally invariant theories.
These theories may describe many bose
fields $\chi$ interacting  with  the inflaton field $\phi$ with different coupling constants:
\begin{eqnarray}
{\cal L} &=& - {M_p^2\over 16 \pi}R +  {1 \over 2} \phi_{,i} \phi^{,i} -{\lambda\over 4} \phi^4
\ \nonumber\\
& & + {1 \over 2}\sum_m  \chi_{{m,i}} {\chi_m}^{,i} -
 {1 \over 2} g_m^2 \phi^2 \chi_{m}^2
-  {1 \over 2} \xi_m R  \chi_m^2 \  ,
\label{lagr}
\end{eqnarray}
Here $\chi_m$ stands for the $m$-th scalar field interacting with  the
inflaton field  with the coupling constant $g_m$,
and interacting with curvature $R$ with the coupling constant $\xi_m$.
 The equation for fluctuations in this  general  model
(\ref{fluc3})
unifies the equations for fluctuations in the conformal
models mentioned before
 including $\delta \phi$ fluctuations.

Strictly speaking, this model is conformally invariant only for a specific choice of the parameters $\xi_m$: \, $\xi_m = {1\over 6}$. Nevertheless,  in this paper we will consider the simplest models with $\xi_m = 0$. As we   will see shortly, this difference is not going to be very important because the average value of $R$ vanishes when $\phi \ll M_p$.

We will see that for the conformally invariant theories the only parameter   actually responsible for the structure of the
resonance for the field $\chi_m$ is the ratio
${{g^2_m} \over \lambda}$.
Furthermore, we will find that the strength of the resonance
and the number and widths of the instability bands
for the field   $\chi_m$
in the theory (\ref{lagr})   depends on   ${g^2_m \over \lambda}$   non-monotonically.
To get a general picture,
we will construct the stability/instability chart for the
  equation for fluctuations on the two dimensional plane $(k^2,
 {g^2 \over \lambda})$, see Fig. 3.
The stability/instability chart gives us insight
into the structure of the resonances in the conformally invariant theories.
 From this it will immediately be clear which of the fields
 $\chi_n$ of (\ref{lagr}) will be most amplified
during preheating. The stability/instability chart
unifies our knowledge of the resonance for the various
conformal models thus far considered in the literature.

Note that the class of theories we are going to investigate include in particular the theory ${\lambda\over 4}\bigl(\sum\limits_{i=1}^N \phi^2_i\bigr)^2$ of an $N$-component scalar field $\phi_i$. This theory has $O(N)$ symmetry. One can identify the inflaton field $\phi$ in this theory with the field $\phi_1$. Then the quantum fluctuations of this field, just like the quantum fluctuations in the theory of a one-component field ${\lambda\over 4}\phi^4$, will have effective mass squared $3\lambda \phi^2$, whereas the fluctuations of all other fields will have effective mass squared $\lambda \phi^2$. Therefore, the equation  for the growth of  the fluctuations of the field $\phi = \phi_1$ (neglecting backreaction) will coincide with the  equation for the growth of fluctuations of the field $\chi$ coupled to the field $\phi$ with the coupling constant $g^2 = 3\lambda$. Meanwhile, the equation  for the growth of  the fluctuations of the fields $\phi_i$,\, $i \not = 1$,   will coincide with the  equation for the growth of fluctuations of the field $\chi$   with the coupling constant $g^2 = \lambda$. This regime is especially important in the limit $N \to \infty$, where the main contribution to particle production is given by the modes with $i \not = 1$.
Thus, the cases $g^2 = \lambda$ and $g^2 = 3\lambda$ are especially interesting and deserve careful investigation.

This paper is organized as follows. In Sec. II we will describe the evolution
of the
background inflaton field $\phi(t) $ after inflation in the theory with the effective potential
 $V(\phi)={1 \over 4} \lambda \phi^4 $.
In Sec. II we will give its analytic solution.
Then, in Sec. III, we derive  the
equations for fluctuations of the fields $\chi$ and $\phi$
in the conformally invariant theory, and reduce these to equations
in Minkowski space-time.
We show that these equations can ultimately be reduced to a single
Lame  equation with just one parameter, ${g^2 \over \lambda}$.
In
Sec. IV we
solve the resonance equations numerically for an arbitrary
 ${g^2 \over \lambda}$ and arbitrary momentum, $k$, of fluctuations.
This allows us to produce the main result of our paper; we
construct the stability/instability chart for fluctuations in
the conformally invariant theories.
In Sec. V we discuss the particular ranges and values of the
parameter  ${g^2 \over \lambda}$ where the
 analytic methods for the description of the resonance
can be developed.
In Sec. VI - IX we perform an analytic investigation of the resonance 
for some  particular  values of  ${g^2 \over \lambda}$.
For different values of ${g^2 \over \lambda}$ different
analytic approaches will be developed. We report
a new method to treat the resonance when
${g^2 \over \lambda} = {n(n+1) \over 2}$, where $n$ is an integer.
We show that the solutions for ${g^2 \over \lambda} = {n(n+1) \over 2}$ can be found
in closed, analytic form.  This is done explicitly for the most interesting   
cases, $n=1$ and $n=3$ (i.e. for $g^2 =\lambda$ and $g^2 = 3\lambda$), in Sec. VI, VII, and the Appendix.
We also consider the two opposite limits
${g^2 \over \lambda} \ll 1$ and ${g^2 \over \lambda} \gg 1$
in Sec. VIII and IX respectively.
Sec. X  contains a discussion of the
self-consistent dynamics of the system including backreaction of the
 created particles.
  In Sec. XI  we describe the restructuring of the resonance
which occurs
when the backreaction is incorporated into the equations for fluctuations. We show that this is the leading effect which terminates the resonance in the theory ${\lambda\over 4} \phi^4$.
In Sec. XII we discuss the modifications of the theory of preheating which appear when the inflaton field $\phi$ is massive. This allows us to unify   the results obtained in this paper with the results of our preceding investigation of preheating in the theory of the massive inflaton field \cite{KLS97}. Finally, in Sec. XIII,
we give a summary of our results and discuss their
possible implications.

 \section{\label{EVOLUTION} Evolution of the Inflaton Field}

We consider   chaotic inflation with the potential
 $V(\phi)={1 \over 4}\lambda \phi^4$.
During inflation the leading contribution to the energy-momentum tensor is
given by the inflaton scalar field $\phi$.
The evolution of the (flat) FRW universe is described by the Friedmann
equation
\begin{equation}
H^2= {8\pi \over 3 M_p^2 } \biggl( {1 \over 2}\dot \phi^2 +
{\lambda \phi^4 \over 4} \biggr)\ ,
\label{E0}
\end{equation}
where $H={\dot a / a}$.
Let us note one more useful
relationship between $H(t)$ and $\phi(t)$ which follows
from the Einstein equations
\begin{equation}
\dot H= - {4 \pi   \dot \phi^2\over M_p^2} \ .
\label{E01}
\end{equation}
The equation for the classical field $\phi(t)$ is
\begin{equation}
\ddot \phi + 3H\, \dot \phi + \lambda \phi^3 =0\ .
  \label{KG0}
\end{equation}
For sufficiently large initial values of $\phi > M_p$,
the friction term, $3H \dot \phi$,  in (\ref{KG0})
dominates over $\ddot \phi$ and
the potential term in (\ref{E0}) dominates over the kinetic term. This is the inflationary stage,
 where the universe expands
 \mbox{quasiexponentially}, $a(t)=a_0 \exp \bigl( \int dt H(t) \bigr)$.
 With a decrease of the
field $\phi$ below $M_p$, the ``drag'' term $3H\, \dot \phi$
gradually becomes less
 important and inflation terminates at $\phi  \sim M_p/2$.
After a short stage of fast rolling down, the inflaton field
 rapidly oscillates around the minimum of $V(\phi)$
with the initial amplitude $\Phi_0 \sim 0.1 M_p$.
Although this value is below the magnitude needed for inflation,
it is still very large.

The character of the classical oscillations of the homogeneous scalar field
depends on the shape of its potential $V(\phi)$.
In Ref.  \cite{KLS97} we  considered  the theory with the
 quadratic potential
$V(\phi)={1 \over 2} m \phi^2$. In that theory the
fluctuations are harmonic, $\phi(t) = \Phi(t)\sin{  mt }$,
with the amplitude decreasing like
$\Phi(t) \approx {M_p   \over \sqrt{3\pi} mt}
\propto {a^{- 3/2}}$.
The scale factor at the stage of  oscillations is
$ a(t) \approx a_0 t^{2/3}$, and
the energy density
of the inflaton field  decreases in the same way as the energy density of
nonrelativistic matter $\propto a^{ -  3}$.

\begin{figure}[t]
\centering
\leavevmode\epsfysize=5.3cm \epsfbox{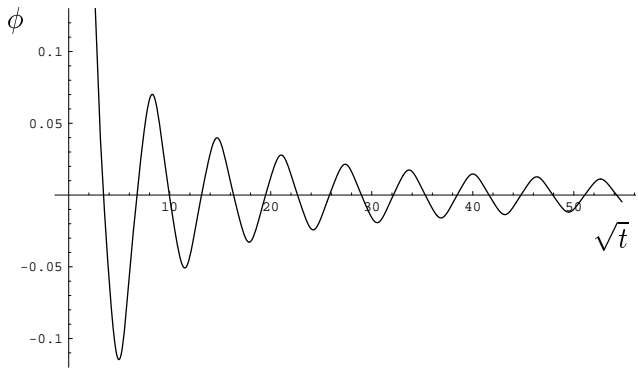}\\

\caption[fig1u]{\label{fig1u} Oscillations of the field $\phi$ after inflation
in the theory ${\lambda\phi^4\over 4}$. The value of the scalar field here and in
all other figures in this paper is measured in units of $M_p$, time is
measured
in units of $(\sqrt\lambda M_p)^{-1}$.}
\end{figure}

In  the theory with the potential  $V(\phi)={1 \over 4} \lambda \phi^4$,
which we consider in this paper,
the inflaton  oscillations  are not sinusoidal. The amplitude $\Phi$ of the oscillations of the scalar field $\phi$
in the limit $t \to \infty$ approaches the asymptotic 
 regime
\begin{equation}
 \Phi(t) \approx  {1\over \sqrt t} \left({3M^2_p \over8\pi \lambda}\right)^{1/4}   \sim~ {M_p \over 10
N}\ ,
\label{870}
\end{equation}
where $N$ is the number of oscillations after the end of inflation.

To make calculations  in this theory, and in particular,
 to find the
form of the  oscillations, it is  convenient
to make a conformal transformation of the space-time metric
and the fields. For this we use
 the conformal time
\begin{equation}\label{ct}
\eta = \int {dt\over a(t)}\ ,
\end{equation} and the conformal field, 
\begin{equation}\label{cf}
\varphi=a \phi .
\end{equation} 
In the  coordinates $(\eta, {\bf x})$
the Klein-Gordon equation (\ref{KG0}) for $\varphi$ is
\begin{equation}
\varphi''  + \lambda \varphi^3 - {a '' \over a}\varphi =0\ ,
\label{KG1a}
\end{equation}
where $'$ stands for the derivative with respect to
the conformal time, ${d \over d \eta}$.
The Friedmann equation (\ref{E0}) in these variables is
\begin{equation}
{{ a'^2}  }={{ 8\pi  }  \over 3 M_p^2 }
\biggl( {1 \over 2} \Bigl(\varphi'- \varphi {a'\over a}\Bigr)^2 +
{\lambda \varphi^4 \over 4} \biggr)   \ .
\label{E1a}
\end{equation} 

As one can see from Eq. (\ref{KG1a}), the equation of motion for the field $\varphi$ in the time variable $\eta$ does not look exactly as the equation for the theory ${\lambda\over 4}\varphi^4$ in Minkowski space. In order to achieve it one would need to add the term ${\phi^2\over 12} R$ to the Lagrangian. However, this subtlety is not very important. First of all,  soon after the end   of inflation  one has ${\lambda\over 4} \phi^4 \gg {\phi^2\over 12} R$, and   $\lambda \varphi^3 \gg {a '' \over a}\varphi$. Moreover, it is known that the energy-momentum tensor of the field $\phi$ in the theory  ${\lambda\over 4} \phi^4$, when averaged over several oscillations, is traceless ($p = \rho/3$) \cite{Turner}. In this case one has $R = 0$, $a(\eta) \sim \eta$, and  $ a''= 0$, so that the last term in Eq. (\ref{KG1a}) vanishes:
\begin{equation}
\varphi''  + \lambda \varphi^3  =0\ .
\label{KG1}
\end{equation}

The Friedmann equation (\ref{E1a}) averaged over several oscillations of the field $\phi$ in the regime $\phi \ll  M_p$ also takes a very simple form:
\begin{equation}
{{ a'^2}  }={{ 8\pi  }  \over 3 M_p^2 }
\biggl( {1 \over 2}  \varphi'^2 +
{\lambda \varphi^4 \over 4} \biggr) \equiv { 8\pi   \rho_\varphi  \over 3 M_p^2  } \ ,
\label{E1}
\end{equation}
where we have introduced the conformal energy density, $\rho_\varphi= {1 \over 2} \varphi'^2 +
{\lambda \over 4} \varphi^4$.

It is convenient to express $\rho_\varphi$ in terms of the amplitude $\tilde\varphi$ of the oscillations of the field $\varphi$:~ $\rho_\varphi=  
{\lambda \over 4} \tilde\varphi^4$. 
Equation (\ref{KG1}) has an oscillatory solution with a constant
 amplitude and the  conformal energy  $\rho_\varphi$.
Then from (\ref{E1}) we find  
\begin{equation}
  a(\eta)=\sqrt{2 \pi\lambda     \over 3 }{\tilde\varphi^2\over M_p}\, \eta \ , ~~~~~ t=\sqrt{  \pi \lambda \over 6 }{\tilde\varphi^2\over M_p}\, \eta^2 \ .
\label{scal}
\end{equation}
As we expected, in this regime the  last term ${a '' \over a}\varphi$ in the equation (\ref{KG1}) vanishes.

Eq. (\ref{KG1})  can be reduced to the canonical equation for an
 elliptic function. Indeed, let us use a dimensionless conformal time
variable
\begin{equation}
x \equiv  \sqrt\lambda \tilde\varphi \eta=
\biggl({ 6 \lambda M_p^2 \over \pi   }\biggr)^{1/4} \sqrt{t} \ .
\label{variable}
\end{equation}
Then we can rescale the function
 $\varphi \equiv a \phi =  \tilde\varphi f(x)$.
The function $f(x)$ has an amplitude equal to unity and obeys the
 canonical equation for the elliptic function. The integral of this equation,
$f'^2={1 \over 2} ( 1 - f^4)$, has the solution in terms of an elliptic
cosine
\begin{equation}
f(x) = cn \Bigl(x - x_0,  { 1 \over \sqrt{2}}\Bigr) \ .
\label{elliptic}
\end{equation}
As claimed, oscillations in this theory  are not sinusoidal but
are given by an elliptic function.
The energy density
of the field $\phi$  decreases in the same way as the density of
radiation, i.e. as $ a^{ -  4}$.
 
\begin{figure}[t]
\centering
\leavevmode\epsfysize=5.6cm \epsfbox{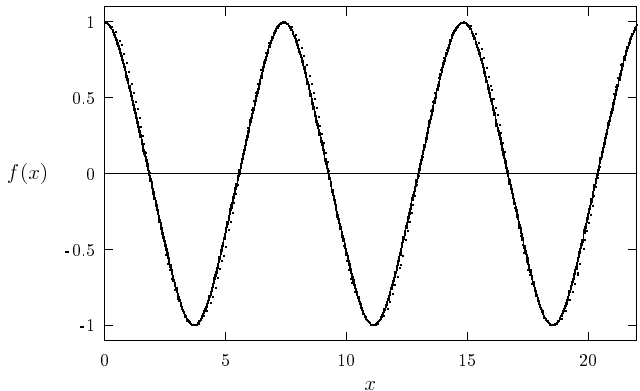}\\
\caption[fig1]{\label{fig1}

The exact solution (\ref{elliptic})
for the oscillations of the inflaton field after inflation
in the conformally invariant theory ${1 \over 4} \lambda \phi^4$.

We show the field in
rescaled conformal field and time
 variables, $f(x)= {\varphi\over\tilde\varphi}$ (solid curve) and the
first term, $ \cos(0.8472x)$,
 in its harmonic expansion  (\ref{series})   (dotted curve).}
\end{figure}

The solution (\ref{elliptic}) has some interesting properties which are not usually elucidated
in the literature.
It matches the solution describing the slow rolling of the field $\varphi$ at the end of  inflation   if one takes $x_0 \approx 2.44$.
The period of the oscillations (in units of $x$) is
 $T=4 K( { 1 \over \sqrt{2}})=
{\Gamma^2(1/4) \over \sqrt{\pi}} \approx 7.416$,
$K$ stands for the complete  elliptic integral of the first kind.
The effective  frequency of oscillations is
$2\pi/T \approx 0.8472$ \cite {KLS}.
The  value of $f^4$ averaged over a period is $ {1 \over 3}$.
The potential
energy density ${1 \over 4}   {\lambda \varphi^4}$ averaged over a period of oscillation is equal to
 ${1 \over 3}\rho_\varphi$, and the average kinetic energy
  ${1 \over 2}  \varphi'^2$ is given by $ {2 \over 3}\rho_\varphi$.

The elliptic cosine can be represented as follows:
\begin{equation}
f(x)={8\pi\sqrt{2} \over T}
\sum_{n=1}^{\infty}{e^{-\pi(n-1/2)} \over {1+e^{-\pi(n-1/2)}}}
\cos {2\pi(2n-1) x \over T} \ .
\label{series}
\end{equation}
The amplitude of the first term in this sum is $0.9550$;
the amplitude of the second term is much smaller, $0.04305$.
The full solution  (\ref{elliptic}) is plotted in Fig. 1 (solid curve),
alongside  the leading harmonic term
in the series (\ref{series}),
 $\cos 0.8472 x $  (dotted curve). 
Although the
first  harmonic term is very close to the
 actual form of oscillations, it will be important
 for the investigation of the general structure of stability/instability bands  in this theory that
$f(x)$ is not exactly  equal to $\cos {2\pi x \over T}$.

 \section{\label{fluctuations}Equations for Quantum Fluctuations of the fields $\phi $ and $\chi$}

We
 will consider here the interaction  between
 the {\it classical }  inflaton field, $\phi$,
and the massless,  {\it quantum} scalar field,
$\hat\chi$, with the Lagrangian (\ref{lagr}).
 The Heisenberg representation of the quantum scalar field $\hat \chi$ is
$$\hat \chi(t, {\bf  x})  =
{1\over{(2\pi)^{3/2}}} \int d^3k\ {\Bigl( \hat a_{{k}} \chi_{
k}(t)\, e^{ -
i{{\bf k}}{{\bf x}}}
+    \hat a_{{k}}^ +    \chi^*_k(t)\, e^{i{{\bf k}}{{\bf x}}}
\Bigr)}  ,
$$
where $\hat a_{{k}}$ and $\hat a_{{k}}^ +   $ are the
annihilation and creation operators. For a flat Friedmann background with
scale factor $a(t)$, the temporal
part of the eigenfunction with comoving momentum ${\bf k}$ obeys the
following equation:
\begin{equation}
\ddot \chi_k  +    3{{\dot a}\over a}\dot \chi_k +   {\left(
{k^2\over a^2}
   +  g^2\phi^2 \right)} \chi_k  = 0 \ .
\label{fluc1}
\end{equation}
As we mentioned in the previous section, at the stage of oscillations when $\phi \ll M_p$ the average value of the curvature  $R$ vanishes,  so one can neglect the term
$\sim \xi\phi^2  R$.

The self-interaction ${1 \over 4} \lambda \phi^4$ also leads to the
generation of fluctuations of the field $\phi$.
The  equation for the  eigenmodes $\phi_{\bf k}(t)$ is
\begin{equation}
\ddot {\phi_k}  +    3{{\dot a}\over a}~{\dot \phi_k}+   {\left(
{k^2\over a^2}
   +  3 \lambda \phi^2 \right)} \phi_k  = 0 \ .
\label{fluc2}
\end{equation}
Note that this equation is identical to equation (\ref{fluc1})
with $g^2 = 3 \lambda$. Therefore, the study of the
fluctuations $\phi_k$ in  the ${1 \over 4} \lambda \phi^4$ model
is a particular case of the general equation
for fluctuations (\ref{fluc1}).

The physical momentum,  ${\bf p} = {\bf k \over a(t)}$,
in equation  (\ref{fluc1}) is redshifted
in the same manner as the background field
amplitude, $\phi(t)={\varphi  \over a(t)}$.
Therefore, the redshifting of momenta can be eliminated
from the evolution of $\chi_k$.
Indeed, let us
use the conformal transformation of the mode function
$ X_k(t)= a(t)\chi_k(t) $ and
 rewrite the mode equation for $X_k(t)$  with the dimensionless
conformal time $x$  (see Eq. (\ref{variable})):
\begin{equation}
X_k''  +  {\left(\kappa^2  +
 {g^2\over \lambda}  cn^2 \Bigl(x,  { 1 \over \sqrt{2}}\Bigr)
 \right)} X_k  = 0 \ ,
\label{fluc3}
\end{equation}
where for simplicity we drop the initial value of $x_0=2.44$.
In this form the equation for fluctuations does not depend on the
expansion of the universe and is completely reduced to the similar
problem in Minkowski space-time.
This is a special feature of the
conformally invariant theory
 ${1 \over 4} \lambda \phi^4  + {1 \over 2} g^2 \phi^2 \chi^2 $.

 For the fluctuations of the field $\varphi = a \phi $ one has
\begin{equation}
\varphi_k''  +  {\left(\kappa^2  +
 3  cn^2 \Bigl(x,  { 1 \over \sqrt{2}}\Bigr)
 \right)}\varphi_k  = 0 \ .
\label{fluc3aa}
\end{equation}

Equation  (\ref{fluc3}) will be the master equation for our
investigation of the resonance in the conformally invariant theory.
The comoving momentum $k$ enters the equation in the combination
\begin{equation}
\kappa^2={ k^2 \over  \lambda   \tilde\varphi^2} \ .
\label{momenta}
\end{equation}
Therefore the natural units of the momenta $k$ is $\sqrt\lambda \tilde\varphi$.
Equation  (\ref{fluc3})  describes
oscillators,  $X_k$,    with a variable frequency
\begin{equation}
\omega^2_k= \kappa^2  +
 {g^2\over \lambda}  cn^2 \biggl(x,  { 1 \over \sqrt{2}}\biggr) \ ,
\label{omega}
\end{equation}
which periodically depends on time, $x$.
It is well known that in this case the solutions  $X_k$
are exponentially unstable:  $X_k(x) \propto e^{ \mu_k x}  $.
If  we choose the vacuum positive-frequency   initial condition,
$ X_k(x) \simeq { e^{-i\kappa x } \over \sqrt{2\kappa} }$,
we then expect the exponentially fast
creation of  $\chi$-particles ($n_k \propto  e^{2 \mu_k x}  $)
 as the inflaton field oscillates.
The strength of interaction with the periodic oscillations
$ cn^2 \bigl(x,  { 1 \over \sqrt{2}}\bigr)$ is given by the
dimensionless coupling parameter $g^2/\lambda$.
This means that the condition of  a  broad parametric resonance
does not require a large initial amplitude of the inflaton field,
$\phi_0$, as in the case of the quadratic potential \cite{KLS}.
As we will see, the combination of parameters  $g^2/\lambda$   ultimately
defines the structure of the parametric resonance in the theory.
It turns out that the strength of the resonance
depends rather non trivially  (non-monotonically)
on this parameter.

 From a mathematical point of view, the mode equation  (\ref{fluc3})
belongs to the class of   Lame equations \cite{erd}.
In the context of preheating this was first noticed in \cite{KLS},
and then thoroughly studied  for  $O(N \to \infty )$
theory (i.e. for $ g^2=\lambda$) in \cite{Boyan96} and \cite{Kaiser97}.
In this paper we perform  a   numerical and
analytical investigation of the parametric amplification of
fluctuations in the conformally invariant theory
 ${1 \over 4} \lambda \phi^4  + {1 \over 2} g^2 \phi^2 \chi^2 $
for an arbitrary parameter ${g^2\over \lambda}$.
In the next section, we present the 
two dimensional  chart of the
stability/instability bands   for the Lame equation  (\ref{fluc3})
in terms of variables $\kappa^2$ and ${g^2\over \lambda}$.
In subsequent sections, we give a new analytic treatment
of the Lame equation in the case $ {g^2\over \lambda}={n(n+1) \over 2}$
with   integer $n$. We will also perform an analytical investigation of the resonance for ${g^2\over \lambda} \ll 1$
and for ${g^2\over \lambda} \gg 1$.

 \section{\label{BROAD} Stability/Instability Chart in the Conformal Theory}

As we shown in the previous section, the equation for vacuum
fluctuations interacting with the inflaton oscillations in the
conformal theories can be reduced to the similar problem in the Minkowski
 space. The equation for fluctuations 
(\ref{fluc3}) in this case contains only two parameters. The first
 parameter is ${{g^2} \over \lambda}$, which    which features the strength of the interaction.
The second parameter is the momentum of vacuum fluctuations 
$\kappa$ in units of the frequency of the inflaton oscillations.
 As is well known, the solutions  $X_k$  of
this equation may be stable or unstable depending on the
particular values for $\kappa$ and ${{g^2} \over \lambda}$ considered.
At the stage of the free resonance when we do not take into account the backreaction of the unstable fluctuations, Eq. (\ref{fluc3})
is an equation with   periodic coefficients, which belongs to  the class of the Lame equations.
The stability/instability chart of another equation with  
 periodic coefficients, the Mathieu equation, is well known and can be found
in many textbooks, see e.g. \cite{AS}.
We are unaware of the  stability/instability charts for the Lame equation,
which describes preheating in the conformally invariant theories.
 Therefore in this section we present 
the stability/instability chart Eq.  (\ref{fluc3}) in variables $\left(\kappa^2,
{{g^2} \over \lambda} \right)$,
which we obtained by solving this equation numerically.

Fig. \ref{fig2a} shows a typical  resonant solution of equation  (\ref{fluc3}).
Though we have plotted the particular case $\kappa^2 = 1.6$,
${{g^2} \over \lambda} = 3$, the form of the resonant solution is generic.
The upper plot demonstrates the amplification of the real part of the
eigenmode $X_k(x)$ (solid curve) in the oscillating inflaton background
(dotted curve).

\begin{figure}[t]
\centering
 \hskip -0.5 cm
\leavevmode\epsfysize= 5.5cm \epsfbox{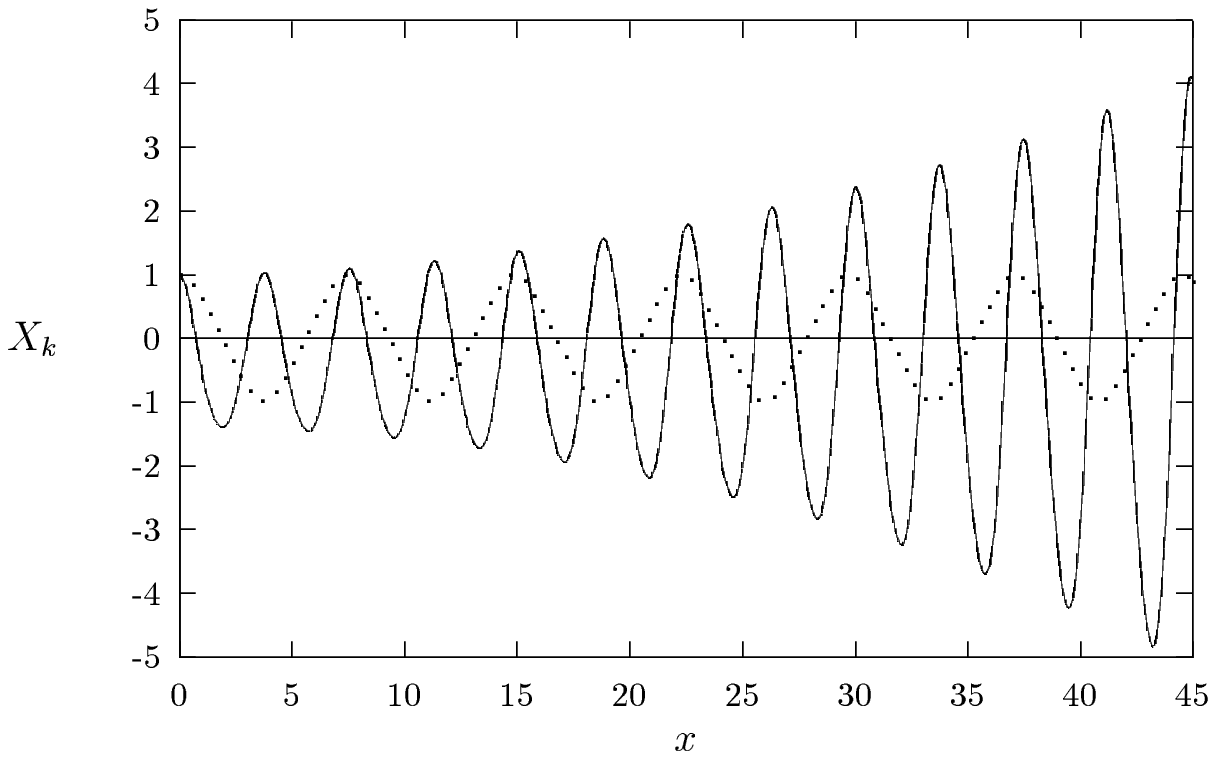}\\
\vskip -0.5 cm
\end{figure}
\begin{figure}[t]
\centering
 \hskip -0.5 cm
\leavevmode\epsfysize= 5.5cm \epsfbox{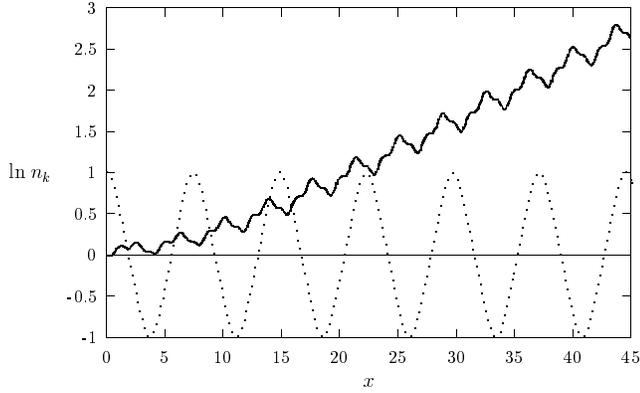}\\
\vskip  0.2cm
 \caption[fig2a]{\label{fig2a}
The typical  resonant production of particles at the
particular choice of
 rescaled comoving momentum $\kappa^2 = 1.6$,
 and the parameter
$\frac{g^2}{\lambda} = 3$.
 The upper plot shows the amplification
of the real part of the eigenmode $X_k(x)$.   The
lower plot shows the logarithm of the comoving particle number density,
$n_k$, calculated with formula (\ref{number}).
 The number of particles grows exponentially,
$\log n_k \approx {2 \mu_k x}$.  In this case, $\mu_k \approx 0.035$. }
\end{figure}
In addition to the investigation of the   rapidly oscillating functions
$X_k(x)$, it is convenient
for analytical and numerical work to consider the
evolution of the comoving number density of created
$\chi$-particles, $n_k$, with comoving momentum $k$.
This can be defined
from the comoving energy density and the energy per particle,
$\omega_k$:

\begin{equation}\label{number}
 n_k={\omega_k\over 2} \left( |X_k|^2 +
{ |\dot X_k|^2 \over \omega_k^2}\right) -{1\over 2}.
\end{equation}

The lower plot of Fig. \ref{fig2a} shows the evolution of the
logarithm of $n_k$ (solid curve) and the inflaton field (dotted curve).
For the  growing solutions  after an initial transitional period the number of
particles increases exponentially, $\ln n_k \approx 2 \mu_k x$,
where $ \mu_k$ is the characteristic exponent of the unstable solution.
 In the
particular case shown, $ \mu_k \approx 0.035$.

\begin{figure}[t]
\centering
 \hskip -0.6 cm
\leavevmode\epsfysize= 6.5cm \epsfbox{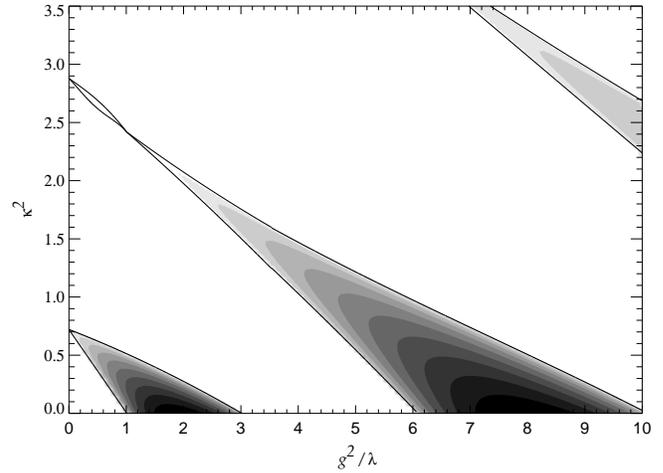}\\
\vskip 0.4cm
 \caption[inst]{\label{inst}
The stability/instability chart for the Lame
 equation for fluctuations  $X_k(x)$
in the variables $(\kappa^2, {g^2 \over \lambda})$,
obtained from  the numerical solution of equation
(\ref{fluc3}).
  Shaded (unshaded)
areas are regions of  instability (stability).
 For instability bands,
the  darker  shade implies
a larger characteristic exponent  $\mu_k$.
Altogether, there are $10$ color steps.
One color step corresponds to
the increment  $\Delta \mu_k=0.0237$,
so the darkest shade corresponds to
maximal  $\mu_k=0.237$, the
least dark shade in the instability bands
corresponds to $\mu_k=0.009$.
For positive $\kappa^2$, there is   only
one instability band  for the
particular values of the parameter $\frac{g^2}{\lambda} = 1$ and $3$.
This  occurs because the higher bands
  shrink  into nodes as   $\frac{g^2}{\lambda}$ approaches
 $ 1$ and   $3$. }
\end{figure}

   For arbitrary values of $\kappa$ and ${{g^2} \over \lambda}$,
we can obtain a numerical solution of equation (\ref{fluc3}) and
exploit the simple relation $\ln n_k \approx 2 \mu_k x$ to extract
the characteristic exponent  for the growing modes.
For the   regions of stability
the characteristic exponent formally is imaginary.
In this way, the stability/instability chart for the Lame equation,
Fig. \ref{inst}, is constructed.  Shaded (unshaded) regions of the
chart indicate values of $\kappa^2$ and ${{g^2} \over \lambda}$ for
which the solutions are unstable (stable).  For the instability bands,
a darker shade indicates a larger characteristic exponent.
 An
immediate result is that, for a given range of ${{g^2} \over \lambda}$,
the largest characteristic exponent will occur for
$\kappa^2 = 0$  between the
integer values ${{g^2} \over \lambda} = {n(n+1) \over 2}$
with  $n$ integer.

   This is   demonstrated in Fig. \ref{mu2fig}, where slices of
the  stability/instability chart show the characteristic exponent
as a function of $\kappa^2$ for various values of ${{g^2} \over \lambda}$.
The top panel of Fig. \ref{mu2fig} plots the cases ${{g^2} \over \lambda} = 1.0, 1.5, 2.0, 2.5,
3.0$, labeled $a$ through $e$ respectively.
${{g^2} \over \lambda} = 1$ corresponds to $n=1$, ${{g^2} \over \lambda} = 3$ corresponds to $n=2$.
As claimed, we see that the largest value of the characteristic exponent
occurs for $\kappa^2 = 0$ at a value of ${{g^2} \over \lambda}$
between the limits $1$ and $3$ (curve $c$).
Similarly, the lower panel of Fig. \ref{mu2fig} plots the cases ${{g^2} \over \lambda}
= 6.0, 7.0, 8.0, 9.0, 10.0$,
labeled $a$ through $e$ respectively. The values 
${{g^2} \over \lambda} = 6$ and $10$ correspond to $n=3$ and $4$.
Again we see that the largest value of the characteristic exponent
occurs for $\kappa^2 = 0$ at a value of ${{g^2} \over \lambda}$
between the limits $6$ and $10$ (curve $c$).

 This stability/instability chart is very similar to the stability/instability chart of the Mathieu equation, but there are important differences as well. For the Mathieu equation there are infinitely many instability bands corresponding to each value of the parameter $q$, which is analogous to our parameter $g^2\over \lambda$. Meanwhile for the Lame equation some of the instability bands may occasionally shrink to a point. As a result, for  ${{g^2} \over \lambda} = 1$ and for   ${{g^2} \over \lambda} = 3$ (Fig. \ref{mu2fig}, curves $a$ and $e$ respectively) there is only one instability band.   This will be shown
analytically in sections VI and VII.  From the
stability/instability chart for the Lame equation, Fig. \ref{inst},
we see that, topologically, this occurs because the higher zones
shrink to nodes as ${{g^2} \over \lambda}$ approaches $1$ and $3$.

\begin{figure}[t]
\centering
 \hskip -0.5 cm
\leavevmode\epsfysize= 5.5cm \epsfbox{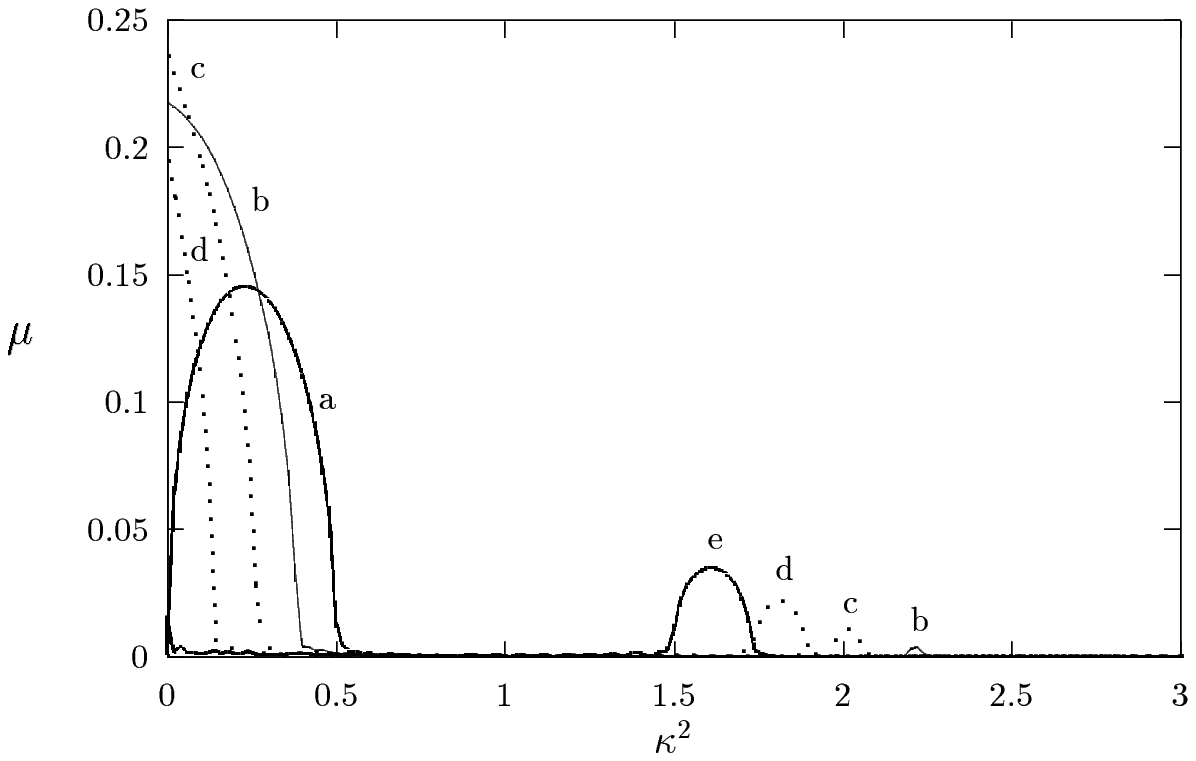}\\
\end{figure}
\vskip -0.5cm
\begin{figure}[t]
\centering
 \hskip -0.5 cm
\leavevmode\epsfysize= 5.5cm \epsfbox{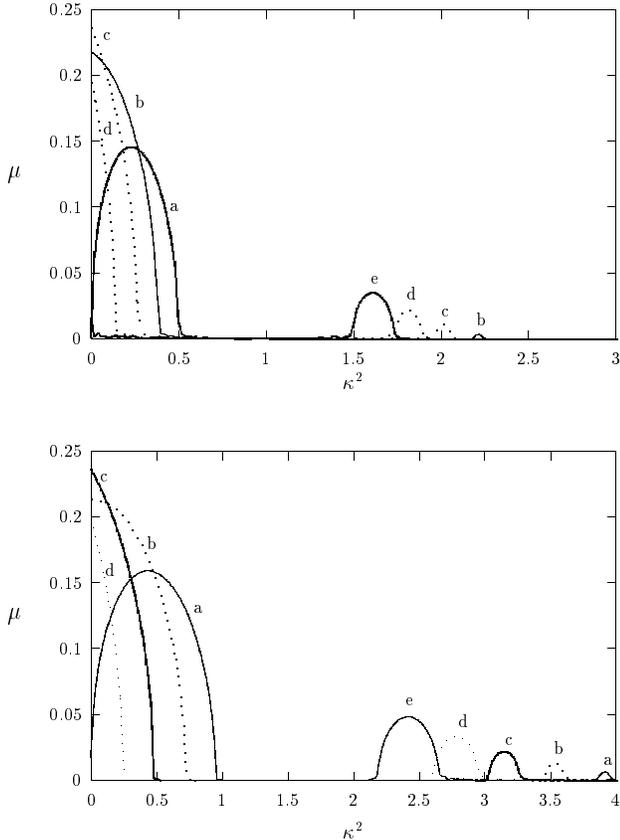}\\
\vskip 0.2cm
 \caption[mu2fig]{\label{mu2fig}
 Slices of the stability/instability chart, Fig.  \ref{inst} ,
reveal the dependence of the characteristic exponent,
$\mu_k$, on $\kappa^2$
for several  particular values of
$\frac{g^2}{\lambda}$. 
For top panel $\frac{g^2}{\lambda}= 1.0$, $1.5$, $2.0$, $2.5$, and $3.0$,
 labeled $a$ through $e$ respectively.
The numerical
curves $a$ and $e$ for $\frac{g^2}{\lambda} = 1$ and
$\frac{g^2}{\lambda} = 3$ are identical to the
 analytic
predictions (\ref{mu2}) of Section VI and (\ref{mu3}) of Section VII. 
For lower panel
$\frac{g^2}{\lambda} = 6.0$, $7.0$, $8.0$, $9.0$, and $10$,
labeled $a$ through $e$ respectively.}
\end{figure}

 	Similarly, there are a finite number of instability bands for 
positive $\kappa^2$ 
whenever $\frac{g^2}{\lambda} = \frac{n(n+1)}{2}$.  However, as for the 
Mathieu equation, all other values of $\frac{g^2}{\lambda}$ have an 
infinite number of instability bands.  This is true in particular for 
$\frac{g^2}{\lambda} \ll 1$ 
where, as we will show in Sec.  VIII, the Lame equation   may be 
formally transformed into the Mathieu equation   with the parameters 
$A \approx 1.3932 \kappa^2$  and $q \approx 0.3464 \frac{g^2}{\lambda} \ll 1$.
Thus, with this change of variables and in the limit $ \frac{g^2}{\lambda} 
\ll 1$, the stability/instability chart for the Lame equation  
is found to coincide exactly with that for the Mathieu equation.

	We now proceed to develop the new analytic results for 
preheating in the theory $\frac{\lambda}{4} \phi^4$, which as we have 
seen, are hinted at by the stability/instability chart for the Lame 
equation, Fig. \ref{inst}.

  \section{\label{lame } Analysis of the  equation for fluctuations}

In this section we begin the analytic investigation of  the Lame equation
 (\ref{fluc3}) for the fluctuations $X_k(x)$.
 In particular, in the next two sections we will
try to find the values of the parameter
${g^2 \over \lambda} $
for which an analytical solutions can be obtain in closed form, and construct these solutions.

We will also investigate the resonance in two limiting cases:
${g^2 \over \lambda} \ll 1$ and ${g^2 \over \lambda} \gg 1$.
In the first case  one can use perturbation theory  in the
small parameter  ${g^2 \over \lambda} \ll 1$, see
 Section VIII \cite{CommSht}. In the opposite limit, ${g^2 \over \lambda} \gg 1$,
we can implement the method of successive
parabolic scattering \cite{KLS97}, see  Section IX.

It is  known that the Lame equation can be solved in terms
of the   transcendental Jacobi functions, which in turn
 are given by series expansions.
Earlier we reported the result  for the characteristic exponent
$\mu=0.0359$ for $\lambda \phi^4$ theory \cite{KLS96,Kof96}.
 Analytic investigation of the resonance using these  transcendental functions
 gives the width of the unstable zone and the maximum of
the characteristic exponent, $\mu_k$, in the physically interesting cases of
the $O(N \to \infty)$ theory
(${g^2 \over \lambda} = 1$  in our convention)
  and the ${1 \over 4 } \lambda \phi^4$ self-interacting theory
(${g^2 \over \lambda} = 3$)
\cite{Boyan96,Kaiser97}.

 However,  calculations involving these transcendental functions are extremely tedious.
Fortunately, it turns out that for  
\begin{equation}
{g^2 \over \lambda} ={n(n+1) \over 2} \ ,
\label{special}
\end{equation}
with $n$ an integer, one can obtain simple, closed-form solutions to
the master equation  (\ref{fluc3}).
This includes in particular the most interesting cases 
$g^2 = \lambda$ and $g^2 = 3\lambda$.

To find the solutions of the  fluctuation equation  (\ref{fluc3})
for ${g^2 \over \lambda} ={n(n+1) \over 2}$, we will rewrite Eq. (\ref{fluc3})  in the so-called algebraic form. We will use the ``time''  variable $z$ instead of $x$:
\begin{equation}
z(x)= cn^2 \Bigl(x,  { 1 \over \sqrt{2}}\Bigr)\ ,~~\
{d \over dx}= \sqrt{2z(1-z^2)}   {d \over dz} \ ,
\label{z}
\end{equation}
Equation  (\ref{special}) for
fluctuations becomes
\begin{equation}
2z(1-z^2) {d^2 X_k \over dz^2}+(1-3z^2) {d X_k \over dz}
+\bigl( \kappa^2 + {g^2 \over \lambda}z\bigr)X_k=0 \ .
\label{fluc4}
\end{equation}
Omitting the lower index $k$ for simplicity,
let  $X_{1}(z)$ and $X_{2}(z)$ be
 two linearly-independent solutions of (\ref{fluc4}). One of them exponentially grows, another exponentially decreases during the resonance.
Let us also introduce  the bilinear combinations
$X_1^2$, $X_2^2$, and $X_1 X_2$. From (\ref{fluc4}) it follows that 
these bilinear combinations obey a third order equation
\begin{eqnarray}
&&2z(z^2-1) {d^3 M \over dz^3}+(9z^2-3){d^2 M \over dz^2}  \nonumber \\ 
&&- 2\left[ \Bigl(  2{g^2 \over \lambda}-3\Bigr)z
     +2 \kappa^2   \right] {d M \over dz}-2{g^2 \over \lambda} M =0 \ .
\label{polin}
\end{eqnarray}
The three solutions, $M(z)$, of this equation  correspond
to the three bilinear combinations of $X_1$ and $X_2$.
The crucial observation is that for   
 ${g^2 \over \lambda} ={n(n+1) \over 2}$ 
the equation
(\ref{polin}) admits a polynomial solution of degree $n$.
 In the particular cases $n=1$ and $n=2$, we have
\begin{eqnarray}
&&n=1:\ ~~ M_1(z)= z-2  \kappa^2   \ ,    \nonumber\\
&&n=2:\ ~~ M_2(z)=z^2 -{ 2 \over 3} \kappa^2 z-1 +{4 \over 9}  \kappa^4 \ .
\label{polinom}
\end{eqnarray}
Obviously, the polynomial function $M(z)$ must be the product of an exponentially growing solution and an exponentially decreasing one, i.e. $M(z)= X_1(z) X_2(z)$ in the resonance zone.
From this, as we will show in the next two sections,  one can construct the
 closed-form solutions $X(z)$.

Therefore, in the physically interesting cases $n=1$ and $n=2$
we will obtain simple closed form  solutions
instead of  the complicated transcendental functions.
This significantly simplifies the
study of preheating in these cases. In particular, we will find
the form of the characteristic exponent $\mu_k$ as a function of $\kappa^2$
in each case.

  \section{\label{n=1} Closed form solution  for}
\vskip -0.9cm
\centerline    {\hskip 7.7 cm ${g^2\over \lambda}= 1$}  
\vskip 0.5cm

In the case $g^2 = \lambda$ equation (\ref{polin}) in the  resonance band   gives
\begin{equation}
X_1(z)X_2(z) = M_1(z) \ ,
\label{n1}
\end{equation}
where
\begin{equation}
 M_1(z)= z-2  \kappa^2 \ .
\label{m1}
\end{equation}
The Wronskian of equation  (\ref{fluc4})
for $X(z)$ is
\begin{equation}
X_1  {d X_2 \over dz}-X_2  {d X_1 \over dz}={ C \over \sqrt{z(1-z^2)}} \ ,
\label{W}
\end{equation}
where $C$ is some constant,  $C=C_1$, to be defined.
From  (\ref{n1}) and   (\ref{W}) we immediately obtain the closed
form solutions
\begin{equation}
X_{1,2}(z) =
\sqrt{ \vert M_1(z)  \vert}
 \exp \left( \pm {C_1 \over 2} \int { dz \over {\sqrt{z(1-z^2)} M_1(z)
 }} \right) \ .
\label{form1}
\end{equation}
Now, substituting this solution back into equation  (\ref{fluc4})
for   $X(z)$, we find the constant $C_1$:
\begin{equation}
C_1= \sqrt{2\kappa^2(1-4\kappa^4)} \ .
\label{C1}
\end{equation}
For exponentially growing solutions, $C_1$ must be real; therefore the exponentially growing solutions for fluctuations
with $\kappa^2 > 0$
take place in a {single}   instability band for which
\begin{equation}
0 <  \kappa^2 <  {1 \over 2} \ .
\label{instab}
\end{equation}
The growing  solution of (\ref{fluc3})   has the form
 $X(x)= e^{\mu_k x} P[z(x)]$, where $P[z(x)]$ is a periodic function
  of the
conformal time $x$. Using (\ref{form1}), we can now find the
characteristic exponent  $\mu_k $
 as a function of $\kappa$. The technical details can be found
in the Appendix.

The final answer is
\begin{equation}
 \mu_k (\kappa) ={2 \over T} \sqrt{2\kappa^2(1-4\kappa^4)} I(\kappa) \ ,
\label{mu2}
\end{equation}
where an auxiliary  function  $ I(\kappa)$ is
\begin{equation}
I(\kappa)= \int\limits_0^{\pi/2} d \theta
{ {\sin^{1/2} \theta} \over {1+2\kappa^2 \sin \theta }} \ .
\label{I}
\end{equation}
Recall that $T \approx 7.416$.
Eq.  (\ref{mu2}) is one of the most
 important analytic results of our paper.
Some numerical values of $\mu_k$ as function of $\kappa^2$
for ${g^2 \over \lambda} =1$
calculated with (\ref{mu2}) are listed in the upper half of the table below.
The analytic form (\ref{mu2}) is in excellent agreement with the
numerical results for this case  plotted in the top panel of Fig. \ref{mu2fig} as  curve $a$.
The maximum value of the characteristic exponent
 for ${g^2 \over \lambda} =1$
is $\mu_{max} \approx 0.1470$  at $\kappa^2 \approx 0.228$,
in agreement with the numerical value for  $\mu_{max}$   of
 Fig. \ref{fig5}.

\vskip 0.5cm

\begin{center}
 \begin{tabular}{|c|c|c|}
\hline
{}~~${g^2/\lambda}$ ~~& ~~$\kappa^2$ ~~& ~~$\mu_k$~~   \\
\hline\hline
{}~~$~1$~ & ~~$~0.0  $~~~ &~~~ $0.000$~~ \\
\hline
{}~~$~1$~ &~~$~0.1 $~~~&~~~$0.1238$~~ \\
\hline
{}~~$~1$~ &~~$~0.2 $~~~&~~~$0.1460$~~ \\
\hline
{}~~$~1$~ &~~$~0.21 $~~~&~~~$0.1466$~~ \\
\hline
{}~~$~1$~ &~~$~0.22 $~~~&~~~$0.1469$~~ \\
\hline
{}~~$~1$~ &~~$~0.228 $~~~&~~~$0.1470$~~ \\
\hline
{}~~$~1$~ & ~~$~0.23 $~~~&~~~$0.1470$~~ \\
\hline
{}~~$~1$~ & ~~$~0.24 $~~~&~~~$0.1468$~~ \\
\hline
{}~~$~1$~ & ~~$~0.25 $~~~&~~~$0.1465$~~ \\
\hline
{}~~$~1$~ & ~~$~0.3 $~~~&~~~$0.1411$~~ \\
\hline
{}~~$~1$~ & ~~$~0.4 $~~~&~~~$0.1117$~~ \\
\hline
{}~~$~1$~ & ~~$~0.5 $~~~&~~~$0.000$~~ \\
\hline
{}~~$~3$~ & ~~$~1.5 $~~~&~~~$0.000$~~ \\
\hline
{}~~$~3$~ & ~~$~1.55 $~~~&~~~$0.02981$~~ \\
\hline
{}~~$~3$~ & ~~$~1.60 $~~~&~~~$0.03570$~~ \\
\hline
{}~~$~3$~ & ~~$~1.61 $~~~&~~~$0.03595$~~ \\
\hline
{}~~$~3$~ & ~~$~1.615 $~~~&~~~$0.03598$~~ \\
\hline
{}~~$~3$~ & ~~$~1.62 $~~~&~~~$0.03594$~~ \\
\hline
{}~~$~3$~ & ~~$~1.625 $~~~&~~~$0.03583$~~ \\
\hline
{}~~$~3$~ & ~~$~1.65 $~~~&~~~$0.03427$~~ \\
\hline
{}~~$~3$~ & ~~$~1.70 $~~~&~~~$0.02460$~~ \\
\hline
{}~~$~3$~ & ~~$~1.732 $~~~&~~~$0.00$~~ \\
\hline
\end{tabular}\\

\end{center}
\vskip 0.3cm

\section{\label{n=3} Closed form solution  for}
\vskip -0.9cm
\centerline    {\hskip 7.8 cm ${g^2\over \lambda}= 3$}  
\vskip 0.5cm

The method of   obtaining a closed-form analytic solution,
$X_k(z)$,  in the case ${g^2 =3 \lambda}$ is similar
to that of the previous section.
In the resonance zone with ${g^2=3 \lambda}$,
equation (\ref{polin}) gives
\begin{equation}
X_1(z)X_2(z) = M_2(z) \ ,
\label{n2}
\end{equation}
where now
\begin{equation}
 M_2(z)=z^2 -{ 2 \over 3} \kappa^2 z-1 +{4 \over 9}  \kappa^2
\label{m2}
\end{equation}
 The Wronskian of equation  (\ref{fluc4})
is the same as  in (\ref{W}),
but with a new constant,  $C= C_2$,  to be defined for this case.
Therefore,  the closed
form solutions are the same as in (\ref{form1}), but
with $ M_2(z)$ in place of $ M_1(z)$.
Substituting this solution  into equation  (\ref{fluc4}),
 we find the constant $C_2$ in this case
\begin{equation}
C_2= \sqrt{ {32 \over 81}\kappa^2
\Bigl(\kappa^4 - {9 \over 4}  \Bigr)(3- \kappa^4 ) \ } \ .
\label{C2}
\end{equation}
Therefore, in the case ${g^2 \over \lambda} = 3$
for $\kappa^2 > 0$,
 there is also  only a {single}   instability band
corresponding to
\begin{equation}
  {3 \over 2}     <  \kappa^2 <  \sqrt{3} \ .
\label{inst2}
\end{equation}

For illustration, we plot the resonant solution $X_k(x)$
in the top panel of Fig.   \ref{fig2a}. Notice that $X_k(x)$
oscillates twice within one inflaton oscillation.
Using the solution (\ref{form1}) with
$ M_2(z)$ and $C_2$, we can find $\mu_k$ in this case;
see the Appendix for details.

The resulting characteristic exponent for ${g^2 \over \lambda} = 3$ is
\begin{equation}
 \mu_k ={8\sqrt{2} \over 9 T} \sqrt{ \kappa^2
\Bigl(\kappa^4 - {9 \over 4}  \Bigr)(3- \kappa^4 )  }\, 
 J(\kappa) \ ,
\label{mu3}
\end{equation}
where the auxiliary  function   $ J(\kappa)$ is
\begin{equation}
J(\kappa)= \int\limits_0^{\pi/2} d \theta
{ {\sin^{3/2} \theta} \over {1+{2\over 3}\kappa^2 \sin \theta
+\big( {4\over 9}\kappa^4-1  \bigr)  \sin^2 \theta    }} \
\label{J}
\end{equation}
 in this case.
Formula  (\ref{mu3})  is another
 important  result of our paper.
Some numerical values of $\mu_k$ as a function of $\kappa^2$
for ${g^2 \over \lambda} =3$
calculated with (\ref{mu3}) are listed in the lower half of the last table.
The analytic form (\ref{mu3}) is in agreement with the
numerical results for this case  plotted in the top panel of Fig. \ref{mu2fig} as   curve $e$.
The maximum value of the characteristic exponent
 for ${g^2 \over \lambda} =3$
is $\mu_{max} \approx 0.03598$  at $\kappa^2 \approx 1.615$,
in agreement with the numerical value for  $\mu_{max}$   of
 Fig. \ref{fig5}.

 \section{\label{limit1} Solution for} \vskip -0.9 cm \hskip 6 cm ${{  g}^2 \over \lambda} \ll 1$ 
\vskip   0.5 cm

In this section we investigate the equation for fluctuations
(\ref{fluc3}) in the limiting case ${g^2 \over \lambda} \ll 1$.
Let us recall that  $f(x)$ is given by the series (\ref{series}), and
hence, $f^2(x)$ in equation (\ref{fluc3}) can be decomposed as
\begin{equation}
f^2(x) = F_0+ F_1 \cos \Bigl({4\pi x \over T} \Bigr)
+F_2 \cos \Bigl({8\pi x \over T}\Bigr)+ ...  \ ,
\label{ser}
\end{equation}
where $F_0=0.4570$,  $F_1=0.4973$,  $F_2=0.04290$ and so on,
but $\sum_{k=0}^{\infty} F_k =1$.
 One can seek $X_k(x)$ in the form of a harmonic series
of terms $ \cos \left({2n\pi x \over T}  \right)$ with slowly
varying coefficients.
If    ${g^2 \over \lambda}$ is a small
parameter,  one can develop an iterative
solution   with respect to ${g^2 \over \lambda}$.
 It is easy to show that the leading contribution
to $X_k(x)$ comes from the lower harmonic:
$ \cos \left({4\pi x \over T}  \right)$. Keeping only this term,
the equation for
 $X_k(x)$ can be reduced to the Mathieu equation
\begin{equation}
{d^2 X_k \over d\tau^2}+ \left( A + 2q \cos 2\tau \right)  X_k =0 \ ,
\label{mathieu}
\end{equation}
where $\tau={2\pi x \over T}$, $A=\left({ T \kappa \over 2\pi}\right)^2$,
and $q={g^2 \over 2\lambda}\left({ T \over 2\pi}\right)^2 F_1$.
Thus, our theory is effectively reduced to
the Mathieu equation  only in the limit
$q \ll 1$, where it has  instabilities in very narrow resonant
bands around $\kappa^2 ={2\pi m \over T}$, \mbox{$m=1$, $2$, \ldots}.
The results of the numerical investigation of the instability zones plotted in
Fig. \ref{inst} indeed show that for ${g^2 \over \lambda} \ll 1$
the parametric resonance corresponds to that of the
Mathieu equation.

\begin{figure}[t]
\centering
 \hskip -0.5 cm
\leavevmode\epsfysize= 5.4cm \epsfbox{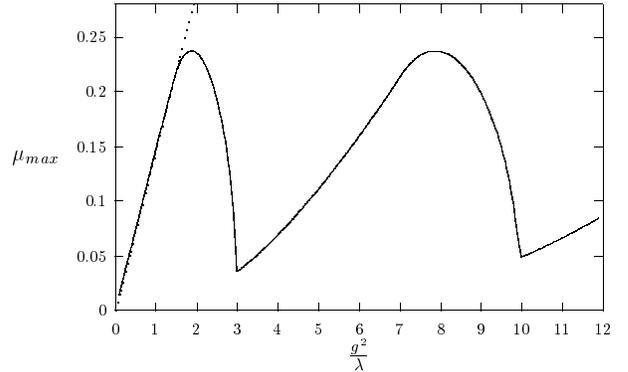}\\
\vskip 0.2cm
 \caption[fig5]{\label{fig5}
The maximum value of the characteristic exponent $\mu_{max}$
extracted from the stability/instability chart, Fig.  \ref{inst},
is plotted as a
function of $\frac{g^2}{\lambda}$ (solid curve).
The function  $\mu_{max}(\frac{g^2}{\lambda} )$ is non-monotonic.
The universal upper limit of $\mu_{max}$ is $0.2377$.
The local minima of the function are gradually increasing with
$\frac{g^2}{\lambda}$, and asymptotically approach  $0.2377$.
The dotted line is
  the prediction $\mu_{max} \approx 0.1467\frac{g^2}{\lambda}$
for $\frac{g^2}{\lambda} \ll 1$, when the mode equation
(\ref{fluc3}) is effectively reduced to  the
Mathieu equation (\ref{mathieu}). }
\end{figure}

The exponentially growing solution  of the Mathieu equation,  $X_k(x) \propto e^{\mu_k x }$,
has a maximum characteristic exponent (in the first zone)
\begin{equation}
\mu_{max} = {g^2 \over 4\lambda}\left({ T \over 2\pi}\right)^2 F_1
 \approx 0.1467  {g^2 \over \lambda} \ .
\label{mu1}
\end{equation}
In Fig. \ref{fig5} we plot the maximum value of the
characteristic exponent as a function of ${g^2 \over \lambda}$
together with the prediction (\ref{mu1})
 for  $\mu_{max} $     from the
 Mathieu equation.
As one can see from Fig. \ref{fig5}, Eq.  (\ref{mu1}) works extremely well even up to
 ${g^2 \over \lambda} \simeq 1$.

 \

 \section{\label{limit2} Analytic Solution for} \vskip -0.9 cm \hskip 7 cm ${{  g}^2 \over \lambda} \gg 1$ 
\vskip   0.5 cm

In this section we consider the limiting case when the parameter
$ {g^2 \over \lambda} $ is very large.
In Fig.  \ref{adiab}  we plot the time evolution of fluctuations
$X_k(x)$ is this case.
 \begin{figure}[t]
\centering
 \hskip -0.4 cm
\leavevmode\epsfysize= 5.5cm \epsfbox{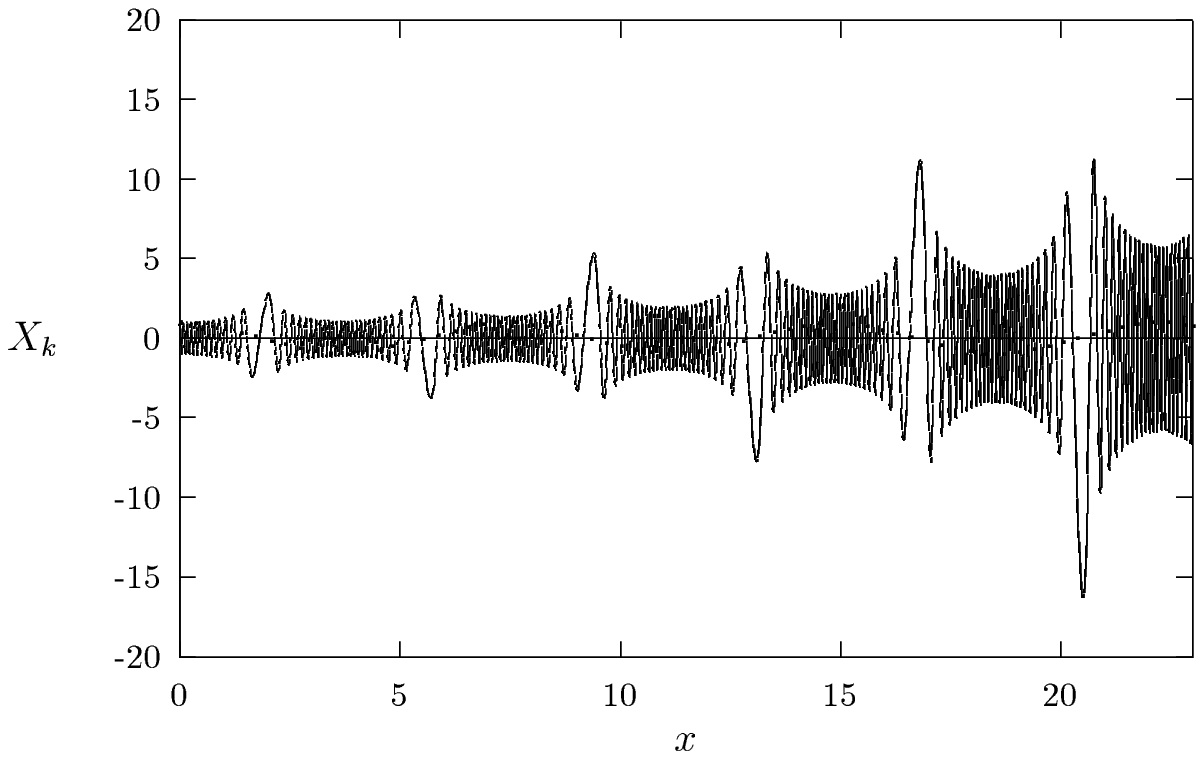}\\
\vskip -0.5cm
\end{figure}
\begin{figure}[t]
\centering
 \hskip -0.4 cm
\leavevmode\epsfysize= 5.5cm \epsfbox{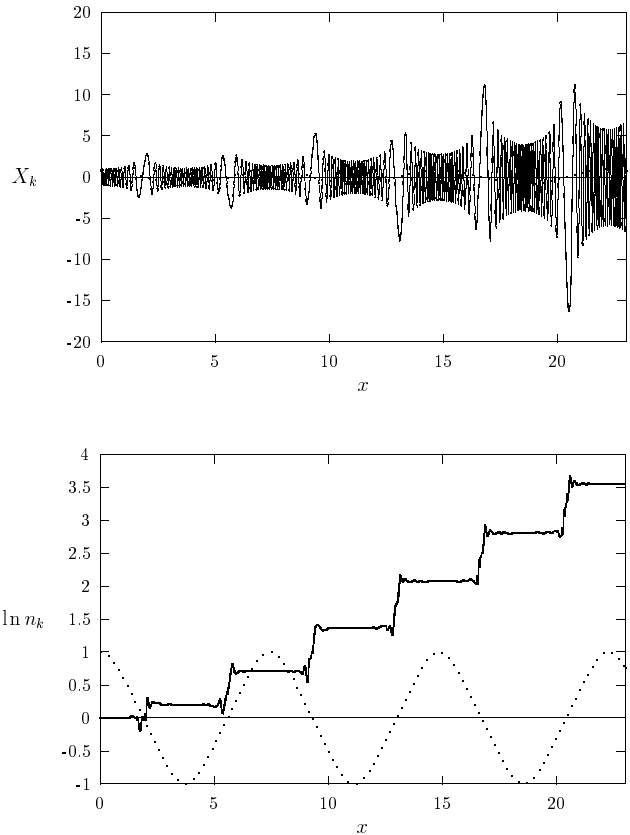}\\
\vskip 0.2cm
 \caption[adiab]{\label{adiab}

The same as in Fig.  \ref{fig2a}  but
for a large value $\frac{g^2}{\lambda} \gg 1$,
here for the particular choice
  $\frac{g^2}{\lambda} = 5050$ and $\kappa^2 = 29.0$.
  The upper plot shows the time-dependence
of the real part of the eigenmode $X_k(x)$,
which demonstrates the adiabatic (semiclassical)
behavior between zeros of the inflaton oscillations (dotted line),
where the comoving occupation number  $n_k$  of
 created particles is constant (lower plot).
 The
lower plot shows $\log n_k$ as a function of time  $x$.
Particle creation occurs in a step-like manner only
in the vicinity of the zeros of the inflaton field,
where the adiabaticity is broken.
The envelope of $\log n_k$ is approximated by
 ${2 \mu_k x}$.
The  characteristic exponent for this example is $\mu_k \approx 0.1$. }
\end{figure}

In Fig.  \ref{adiab}  we plot the number of particles  $n_k(x)$  in a
given mode as a function of time  $x$
calculated from $X_k(x)$ with Eq.  (\ref{number}).  The basic observation is that, for ${g^2 \over \lambda} \gg 1$,
the evolution of the modes  $X_k(x)$ is adiabatic
and the number of particles $n_k(x)$ is constant
between the zeros of the background field.
Changes in the number density of particles occur
only near times $x=x_j$ when the amplitude of the inflaton field
crosses zero, i.e. $\varphi(x=x_j)=0$.
To describe  the effect of a single kick at $x=x_j$,
it is enough to consider the evolution of $X_k(x)$
in the interval when  $\varphi^2(x)$ is small and can therefore    be represented by
its quadratic part $\propto (x-x_j)^2$. This process looks like
wave propagation in a parabolic potential.
Outside of these time intervals,  $X_k(x)$ has a simple,
\mbox{semiclassical} (adiabatic) form.
We can combine the action of the  subsequent
parabolic potentials to find
the net effect of particle creation.
This method of successive parabolic scattering
was formulated and applied to the broad parametric resonance
for the quadratic inflaton potential in \cite{KLS97}.
This method, as we see, can also be applied
to the conformally invariant theory for  ${g^2 \over \lambda} \gg 1$.

We expect that the semiclassical  solution
 is valid everywhere but around
 $x_j$. Thus, prior to scattering at $x_j$, the mode function $X_k(x)$
has the adiabatic form
\begin{equation}
X_k^{j}(x)  ={{\alpha_k^{j}} \over  \sqrt{2\omega_k}}\,
 e^{-i\int_0^x \omega_k dx} +
{{\beta_k^{j}} \over  \sqrt{2\omega_k}}\, e^{+i\int_0^x  \omega_k dx} \ ,
\label{WKB1}
\end{equation}
where the coefficients $\alpha_k^{j}$ and $\beta_k^{j}$
are constant for  $x_{j-1} < x < x_j$ and  normalization yields
$\vert \alpha_k\vert^2- \vert \beta_k\vert^2=1$.
After scattering when $x=x_j$, $X_k(x)$
in the interval
  $x_j < x < x_{j+1}$ again has the adiabatic
form of equation (\ref{WKB1}) but with new constant coefficients,
$\alpha_k^{j+1}$ and $\beta_k^{j+1}$.

The  form  is  essentially  the
 asymptotic  expression
of the  incoming  waves (for  $x < x_j$) and similarly for
 the   the outgoing   waves (for  $x > x_j$) scattered
from a parabolic potential $(x-x_j)^2$
at the moment $x_j$.
Therefore,  the outgoing amplitudes,
 $\alpha_k^{j+1} $ and  $ \beta_k^{j+1} $,   can be expressed in terms of
the  incoming
amplitudes,
$\alpha _k^{j}$ and $  \beta_k^{j}$,   with the help of the reflection
  and transmission
amplitudes  for scattering at a parabolic potential \cite{KLS97}.
For this we need the   mode equation
around a single parabolic potential at $x=x_j$. In the vicinity of
$x_j$,
 $ cn\left(x,  { 1 \over \sqrt{2}} \right) \approx (x-x_j)$.
Then equation (\ref{fluc3}) around  $x_j$ is reduced to the simple
equation
\begin{equation}
{d^2 X_k \over dx^2} + \left( \kappa^2   +
       {g^2 \over 2\lambda} (x-x_j)^2 \right) X_k =0 \ .
\label{parab1}
\end{equation}

The mapping of  $\alpha _k^{j}$,  $  \beta_k^{j}$
into  $\alpha _k^{j+1}$,  $  \beta_k^{j+1}$
in terms of parameters in equation (\ref{parab1})
 reads

\begin{equation}\label{matrix1}
\pmatrix{\alpha_k^{j+1}  \cr\cr \beta_k^{j+1}   \cr } =
\pmatrix{ \sqrt{1+e^{-\pi \epsilon^2}} e^{i\zeta_k}   &
   ie^{-{{\pi \over 2} \epsilon^2} +2i\theta_k^{j}}       \cr \cr
  -ie^{-{{\pi \over 2} \epsilon^2} -2i\theta_k^{j}}
  &  \sqrt{1+e^{-\pi \epsilon^2}} e^{-i\zeta_k}       \cr}
\pmatrix{\alpha_k^{j} \cr\cr \beta_k^{j} \cr}
 \   \nonumber
\end{equation}
where $\zeta_k= \arg \Gamma \left({1+i\epsilon^2 \over 2}\right) $
and  \mbox{$\epsilon^2=\sqrt{2\lambda \over g^2} \kappa^2
= {k^2 \over \sqrt{\lambda/2}\tilde\varphi^2 g}$}.
The  phase accumulated by the moment ${x_j}$ is
$\theta_k^{j}=\int\limits_0^{x_j} dx~ \omega_k(x)
=j  \, \theta_k  $, where
  $\theta_k =2 \int\limits_0^{T/4}
dx~\sqrt{\kappa^2 +   {g^2 \over \lambda}f(x)}   $ is the phase
accumulating within half of a period of the inflaton oscillation.

In the regime when a large  number of
particles  have been created,
$n_k^j= \vert  \beta_k^{j}\vert^2 \gg 1$,
we have  $ \vert  \alpha_k^{j}\vert \approx \vert  \beta_k^{j}\vert$,
so  $\alpha_k^{j}$ and   $  \beta_k^{j}$ are
distinguished by their phases only.
In this case  there is a simple solution of the
matrix equation (\ref{matrix1}):
\begin{equation}\label{a1}
\alpha_k^{j}= { 1 \over \sqrt{2}}\cdot e^{( \mu_k{T \over 2}
 + i\theta_k)\cdot j },
\beta_k^{j}= { 1 \over \sqrt{2}} e^{i\vartheta}
 \cdot e^{( \mu_k {T \over 2} - i\theta_k)\cdot j} \ ,
\end{equation}
where ${\vartheta}$ is a constant phase and $\mu_k$
is the  characteristic exponent. $T \approx 7.416$
is the period of oscillations of the inflaton field
in the variable $x$,
so the number of  particles grows as $e^{2\mu_k x}$.
 Another solution
 is similar to (\ref{a1}) but with the substitution
$\theta_k \to \theta_k+\pi$.

Substituting the solution  (\ref{a1}) into
equation  (\ref{matrix1}), we get an
  equation for the parameters $\mu_k$ and $\theta_k$
\begin{eqnarray}\label{cond4}
 e^{  \mu_k {T \over 2 } }&=&| \cos (\theta_k- \zeta_k)|\sqrt{1+e^{-\pi
\epsilon^2}}\nonumber\\
&+&\sqrt{(1+e^{-\pi \epsilon^2}) \cos^2 (\theta_k- \zeta_k) -1    } \ .
\end{eqnarray}
In the instability zones, the parameter  $\mu_k$
of equation (\ref{cond4}) should be real.
From this we obtain the
  condition
\begin{equation}\label{cond3}
| \tan (\theta_k- \zeta_k)| \leq e^{-{{\pi \over 2} \epsilon^2}} \ .
\end{equation}
for the momentum $k$ to be in a resonance band.

To further analyze the conditions for the strength (\ref{cond3})
 and   widths (\ref{cond4})  of the resonance,
one should calculate  the phase $\theta_k-\zeta_k$. For
${g^2 \over \lambda} \gg 1$ we have
\begin{eqnarray}\label{phase}
\theta_k-\zeta_k &=&2 \int\limits_0^{T/4}
dx~\sqrt{\kappa^2 +   {g^2 \over \lambda}f^2(x)}- \arg
 \Gamma \left({1+i\epsilon^2 \over 2}\right)
\nonumber  \\
&\approx& \pi \sqrt{g^2 \over 2 \lambda}+
\kappa^2\sqrt{\lambda \over 8 g^2} \ln {g^2 \over \lambda}  \ .
\end{eqnarray}

Using equations
(\ref{phase}), (\ref{cond3}), and (\ref{cond4}), we find
the characteristics of the resonance in the regime
${g^2 \over \lambda} \gg 1$. From (\ref{cond3}) it follows that
the resonance is efficient for $\epsilon^2 \leq \pi^{-1}$, i.e for
\begin{equation}\label{typ}
\kappa^2 \leq \sqrt{ g^2 \over 2\pi^2\lambda} \ .
\end{equation}
Equation    (\ref{cond3})    transparently
shows that, for a given ${g^2 \over \lambda}$, there will be a
sequence of stability/instability bands as a function of $\kappa$. The width  of an instability band, where the resonance occurs,  is $\Delta \kappa^2 \simeq \sqrt{g^2 \over 2\lambda}$.
 Let the integer part of the large number  $\sqrt{g^2 \over 2\lambda}$
be $l$. From (\ref{phase}) it follows that if we   vary $\kappa^2$ within the range
$  2 \pi^2  \sqrt{ g^2 \over 2\pi^2\lambda}
 \bigl({\ln {g^2 \over \lambda}}\bigr)^{-1}$,
then within this interval of $\kappa^2$  the phase
$\theta_k-\zeta_k$ reaches either $l \pi$ or $(l+ 1) \pi$.  
  Then within this resonance band we
get the maximum value $\mu_{max}$ defined by the equation (\ref{cond4})
with $|\cos(\theta_k-\zeta_k)|=1$:
\begin{equation}\label{max4}
 e^{  {T \over 2} \mu_{max}  }=\sqrt{1+e^{-\pi
\epsilon^2}}+e^{-{\pi \epsilon^2 \over 2}} \ .
\end{equation}
The characteristic exponent  $\mu_{max}$ is a non-monotonic function of
 ${g^2 \over \lambda}$. If the value of the
parameter ${g^2 \over \lambda}$ is
exactly equal to $2l^2$ where $l$ is an integer,
then the strongest resonance occurs at $\kappa^2=0$,\footnote{It is easy to see from (\ref{phase})  that
the mode  $\kappa^2=0$ is within the resonance band
if $2l^2-l < {g^2 \over \lambda} < 2l^2 +l$.}
 and from
(\ref{max4}) we get
\begin{equation}\label{max5}
\mu_{max}= { 2 \over T} \ln \bigl(1+ \sqrt{2}\bigr) \approx 0.2377 \ .
\end{equation}
This is actually a general result for the upper limit of
$\mu_{max}$ for an arbitrary  ${g^2 \over \lambda}$,
see Fig.  \ref{fig5}.
 If  ${g^2 \over \lambda}$ is not  exactly equal to  $2 l^2$, then
$\mu_{max}$ occurs at a non-zero  $\kappa^2$ and
is smaller than $0.2377$.
It is interesting that in the formal limit
 ${g^2 \over \lambda} \to \infty$ the function
$\mu_{max}\big({g^2 \over \lambda}\bigr)$ asymptotically
approaches the value  $0.2377$ for   arbitrary  ${g^2 \over \lambda}$.
To see this, we have
to check that
 a variation of $\kappa^2 \sim  2 \pi^2
\sqrt{ g^2 \over 2\pi^2\lambda}\,   \bigl(\ln {g^2 \over \lambda}\bigr)^{-1}$
is  compatible with the condition
for an efficient resonance, $\epsilon^2 \leq \pi^{-1}$. This occurs for
${g^2 \over \lambda} \geq e^{2\pi^2} \simeq 10^{8}$.
In Fig.  \ref{fig5}  we see that the  minimal value of $\mu$
as a function of ${g^2 \over \lambda}$ very slowly increases
towards  $0.2377$.
Therefore, although $\mu_{max}$
is not a monotonic function of $ {g^2 \over \lambda} $,\, for ${g^2 \over \lambda} \gg 1$
the resonance is stronger both in terms
of the characteristic  exponent
 $\mu_{max}$ and the width $\kappa^2$.

 \section{\label{recon} Backreaction of created particles}

Thus far we have considered the parametric resonance in the
conformally invariant theory (\ref{lagr}) in an expanding universe
neglecting the backreaction of the amplified fluctuations
of the fields $\phi$ and $\chi$.
In the next two sections we will study the effects related to the backreaction.

In the   theory
 ${1 \over 4} \lambda \phi^4  + {1 \over 2} g^2 \phi^2 \chi^2 $,
the   equation of motion for the inflaton field $\phi(t)$
looks as follows:
\begin{equation}
\ddot \phi + 3H\, \dot \phi +  \lambda \phi^3
+3\lambda \langle   \phi^2\rangle  \phi+
g^2 \langle  \chi^2\rangle  \phi
 =0\ .
\label{KG5}
\end{equation}
The two additional terms are due to the one-loop Hartree diagrams;  $\langle   \phi^2\rangle$ and $\langle  \chi^2\rangle$ stand  for the quantum fluctuations of the fields $\phi$ and $\chi$ respectively.
There are also
higher-loop corrections such as $2g^2 \langle   \phi  \chi^2\rangle$
and $6\lambda \langle     \phi^3\rangle$, which are
not necessarily negligible at the end of preheating
when we may expect $n_{\chi} \sim 1/g^2$ and  $n_{\phi} \sim 1/\lambda$  (or even higher, if one takes into account rescattering).
Here we will work in  the one-loop approximation.

If again we use the conformal transformation  $\eta=\int {dt\over a}$,
$\chi_k={X_k\over a}$ and similarly 
$ \phi_k=\varphi_k/a$, then
 the equation  for  the
background field  $\varphi(\eta)$ is
\begin{equation}
\varphi''  + \lambda \varphi^3
 +3\lambda \langle  \varphi^2\rangle  \varphi+
g^2 \langle  X^2\rangle \varphi  =0 \ ,
\label{KG7}
\end{equation}
with the comoving  vacuum expectation values for  $X$ and
 $\varphi$ correspondingly
\begin{equation}
 \langle X^2 \rangle = {1 \over (2\pi)^3 } \int  d^3k\,
   \vert X_k \vert^2 ,~~\
 \langle   \varphi^2 \rangle
 =  {1 \over (2\pi)^3 } \int   d^3k\,
   \vert  \varphi_k \vert^2.
\label{fluc8}
\end{equation}

The integral of the equation  (\ref{KG7})  coincides with the energy  density
\begin{equation}
\rho_{tot} = {1 \over 2} \varphi'^2 + {\lambda \over 4}\varphi^4 + \rho_\varphi +\rho_{X}  \ .
\label{energy}
\end {equation}
The first two terms describe the energy of the classical field $\varphi$, $ \rho_\varphi$ and  $\rho_{X}$ correspond  to the energy density of $\varphi$-particles and $X$-particles respectively:
 \begin{equation}
\rho_\varphi = {1 \over  (2\pi )^3} \int  d^3k\, \sqrt{k^2+3\lambda \varphi^2}~n_k^\varphi  \ ,
\label{energyxx}
\end {equation}  
\begin{equation}
\rho_{\chi}={1 \over  (2\pi )^3} \int  d^3k\, \sqrt{k^2+g^2\varphi^2}~n_k^X \ .
\label{energy1}
\end{equation}
Here   $n_k^\varphi$ and $n_k^X$ correspond to the occupation numbers of the $\varphi$-particles and $X$-particles.
It is easy to show that $\rho'_{\chi}=g^2\langle \chi^2 \rangle 
\varphi \varphi'$ and $\rho'_{\varphi}=3\lambda\langle \varphi^2 \rangle 
\varphi \varphi'$, and therefore Eq. (\ref{energy}) is an intergal of Eq. (\ref{KG7}).

\begin{figure}[t]
\centering
\leavevmode\epsfysize= 8cm \epsfbox{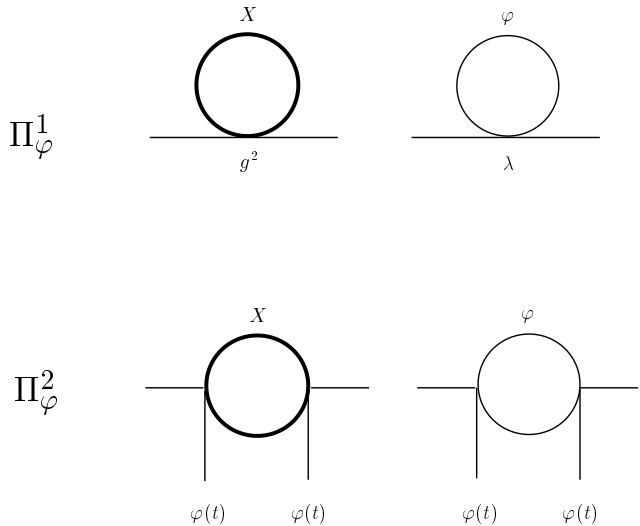}\\
\caption[polariz]{\label{polariz} The diagrams for the polarization operator
of
the field $\varphi_k$. Thin and thick lines represent the
 fields $\phi$ and $\chi$
respectively. Vertical lines correspond to the oscillating background field $\varphi(t)$. $\Pi_\varphi^1$ corresponds to the Hartree approximation
which
takes into account the contribution of $\langle X^2\rangle$ and  $\langle\varphi^2\rangle$.  The
contributions of $\Pi_\varphi^1$ and $\Pi_\varphi^2$ to the effective mass of  $\varphi$-particles
can be comparable
to each other.}
\end{figure}

To close the set of self-consistent equations we need the   equations for the modes $ \varphi_k(x)$ and $X_k$:
\begin{equation}
\varphi_k''(\eta)  +  {\left(k^2  + \Pi_\varphi+
 3 \lambda \varphi^2(\eta)   \right)}
 \varphi_k  = 0 \ ,
\label{fluc10}
\end{equation}
 \begin{equation}
X_k''(\eta)  +  {\left(k^2  + \Pi_{X}+
 {g^2}  \varphi^2(\eta)
 \right)} X_k  = 0 \ .
\label{fluc11}
\end{equation}
The polarization operator $\Pi_\varphi$ consists of
$\Pi^1_\varphi =3\lambda \langle  \varphi^2\rangle  +
g^2 \langle  X^2\rangle$ and
the non-local term $\Pi^2_\varphi$
 which emerges
in the one-loop approximation beyond the Hartree
diagram, see Fig.  \ref{polariz}.

The calculation of the polarization operator  $\Pi^2_\varphi$   in the regime of parametric resonance is rather complicated. Estimates of $\Pi^2_\varphi$ performed in 
  \cite{KLS97} indicate that it can be of the same order
  of magnitude as the
standard Hartree polarization operator $\Pi^1_\varphi$. The polarization operator  $\Pi^2_\varphi$   was not taken into account in the previous treatment 
of  the self-consistent equations for the eigenmodes
 in the ${1 \over N}$ approximation  \cite{Boyan95,Boyan96}, but in fact it may  survive
 in the limit $N \to \infty$ \cite{KLS97}. 

Similarly, the polarization operator  $\Pi_X $ is equal to $g^2 \langle  \varphi^2\rangle$,
 plus an additional non-local term $\Pi_X^2$.
We expect that $\Pi_\varphi \geq 0$, $\Pi_X \geq 0$, as suggested by the Hartree approximation.

A complete calculation of the  polarization operators $\Pi_\varphi$ and $\Pi_X$ is outside the scope of this paper. Fortunately, as we will see in the next section, one need not really know exact expressions for  $\Pi_\varphi$ and $\Pi_X$ in order to make an estimate of the density of produced particles at the time when   the feedback of the amplified fluctuations
  terminates the
parametric resonance.

 \section{\label{term} Dynamical restructuring of the resonance}

In this paper we found that the structure of the
parametric resonance
in terms of its strength and width
 strongly depends on the
parameters of the model.
For example, the parametric resonance
in the simplest conformally invariant theory
 ${1 \over 4} \lambda \phi^4  + {1 \over 2} g^2 \phi^2 \chi^2 $
is very different from that in the  theory
${1 \over 2} m_{\phi} \phi^2  + {1 \over 2} g^2 \phi^2 \chi^2 $
\cite{KLS97}.
In the simplest conformally invariant theories which we consider in this paper the structure of the resonance is determined by 
  the  combination   $g^2/\lambda$.

How does the resonance   develops if the backreaction of the
accumulating fluctuations is taken into account?
The answer to this question also
strongly depends  on the parameter $g^2/\lambda$.

For  illustration we consider
the model of the self-interacting inflaton field
${1 \over 4}\lambda \phi^4$,
 no  $\chi$ field is involved. In this case we shall take $g^2=0$ in all the equations
(\ref{KG7}), (\ref{fluc10}), (\ref{fluc11}). As we already mentioned, if one neglects backreaction, the equations describing the resonance for the modes $\varphi_k$ in this theory coincide with the equations for the modes $\chi_k$ in the theory with $g^2= 3\lambda$. Thus, we can use the results of the investigation of the theory with $g^2= 3\lambda$ obtained in Sec. VII for our analysis.

Historically, the model ${1 \over 4}\lambda \phi^4$ was one of the first models illustrating the general idea of preheating. The investigation of the stability/instability chart for the Lame equation has shown that this model in a certain sense is the least favorable for the development of the resonance: it has only one resonance band, and the characteristic exponent $\mu$ for this theory is anomalously small, see Figs. 4, 6. Originally it was expected that preheating in this model would rapidly transfer about   half of the energy of the oscillating scalar field to the $\phi$-particles, after which the decay of the field $\phi$ would continues  at a much slower pace. However, the results of computer simulations of preheating in this theory indicated that the stage of efficient preheating ends as soon as the fluctuations of produced particles $ { \langle  \varphi^2\rangle}$ grow   to    $0.05\, \tilde\phi^2$ \cite{KhTk961}.  The interpretation of this result, however, was not quite clear. It was conjectured that the resonance terminates because of rescattering of  the $\phi$-particles. It was not   clear also whether the decay of the field $\phi$ continues at a slower pace until this field completely decays, or its decay eventually shuts down.

A complete investigation of this issue is rather difficult. First of all, the theory of rescattering   is not fully developed: various approximations often break down near the end of preheating when   the occupation numbers of particles are anomalously large ($n_k \sim \lambda^{-1}$) \cite{KLS,KLS97}.    Even in the Hartree approximation (or in the $1/N$-approximation) an investigation is extremely complicated \cite{Boyan96,Kaiser97} because it is very difficult to work  with the solutions of equations for the growing modes in terms of the transcendental Jacobi functions. It may be   easier to work with  the solutions   obtained in Sec. VI and VII. We will not perform a complete  investigation of this issue here because, as we argued in the previous section,  one may need to calculate the polarization operator beyond the Hartree approximation, see \cite{KLS97}. Instead,  we will make some simple estimates which will allow us to elucidate   the mechanism which terminates the resonance  in the theory ${1 \over 4}\lambda \phi^4$. 

As we will see,   the main reason for the  termination of the resonance in the theory ${1 \over 4} \lambda \phi^4$ is the restructuring of the resonance band due to the backreaction of created particles. This process   occurs at  $ { \langle   \varphi^2\rangle}\ll \varphi^2$ because the resonance band is very narrow. In the beginning of preheating in the  theory  ${1 \over 4} \lambda \phi^4$   the    instability band is given by the condition ${1.5} \lambda\tilde\varphi_0^2  <  k^2  <  1.73  \lambda\tilde\varphi^2_0 $, where $\tilde\varphi_0$ is the initial amplitude of the oscillations of the field $\varphi$ (\ref{inst2}). It is sufficient  to shift the position of the resonance band in   momentum space by few percent, and the leading resonant modes $\chi_k$ which have been growing since the start of the parametric resonance  will not grow anymore. This will effectively shut down the resonance.

There are two different effects which lead to a restructuring of the resonance band, and these effects act in opposite directions. First of all,  particle production reduces the energy of the scalar field, and therefore reduces the amplitude of its oscillations. This effect tends to reduce the frequency of the oscillations and to move the resonance band towards smaller $k$. On the other hand, the effective mass of the field $\varphi$ grows due to its interaction with the $\phi$-particles. This effect increases the frequency of oscillations and tends to shift the resonance band towards larger $k$.   We will consider here both of these effects.

  To investigate  the decrease of the amplitude of the oscillations due to particle production, one should compare the total energy of the system before and
after the appearance of $\langle\phi^2\rangle$:
\begin{equation}\label{ENERGYb}
{\lambda\over 4}\tilde\varphi^4_0  \approx {\lambda\over 4}\tilde\varphi^4(\eta) + {1\over (2\pi)^3}\int d^3 k \sqrt{k^2+3 \lambda\tilde\varphi^2 }~ n_k   \ .
\end{equation}
Here we calculate   the energy density at the moment when $\varphi' = 0$, and the oscillating field is equal to its amplitude $\tilde\varphi(\eta)$. This amplitude is smaller than 
$\tilde\varphi_0$ due to the transfer of energy to the created particles.

The resonance   is most efficient   in a small vicinity of $k^2 \approx 1.6  \lambda\tilde\varphi^2 $. Therefore, the leading contribution to $\rho_\phi$ is given by integration  near $k^2 = 1.6\lambda
\tilde\varphi^2$:
\begin{equation}\label{ENERGYd}
{\rho_\phi}  \approx    {1\over (2\pi)^3}\int d^3 k \sqrt{4.6 \lambda\tilde\varphi^2}~
n_k  = \sqrt{4.6 \lambda }\, \tilde\varphi \, n_\phi  \ .
\end{equation}
Eqs. (\ref{ENERGYb}) and (\ref{ENERGYd}) give
\begin{equation}\label{ENERGY1}
{\tilde\varphi (\eta) } \approx {\tilde\varphi_0} - {\sqrt{4.6\lambda}\, n_\phi\over \tilde\varphi^2  }  \ ,
\end{equation}

Thus, the creation of $\varphi$-particles  diminishes the  frequency of oscillations, because the frequency of oscillations of the field $\varphi$ in the theory $\lambda\phi^4$ is proportional to its amplitude. To evaluate the significance of this effect one may express it in terms of 
$\langle\phi^2\rangle$ calculated at $\varphi(\eta) = \tilde\varphi$:
\begin{equation}\label{dispersionNEW}
\langle\phi^2\rangle  \approx {1\over (2\pi)^3}\int {d^3 k\  n_k \over \sqrt{k^2+ 3 \lambda \tilde\varphi^2  }} \approx    {n_\phi  \over \sqrt{4.6   \lambda}\,  \tilde\varphi } \  .
\end{equation}
From the last two equations one obtains
\begin{equation}\label{ENERGY1}
{\tilde\varphi^2 (\eta)\over \tilde\varphi_0^2 } \approx  1  - 9.2  { \langle\phi^2\rangle\over \tilde\varphi^2_0  }  \ ,
\end{equation}
which   leads to a proportional shift of the resonance band towards smaller $k^2$. 
This indicates that even a very small amount of fluctuations $\langle\phi^2\rangle \sim 10^{-2} \tilde\varphi^2_0$ may   shift the resonance band away from its original position, which may terminate the resonance for the leading modes $\varphi_k$.

This effect is   partially compensated by the growth of the effective mass of the field $\varphi$.   We will analyse this effect in the Hartree approximation, in which the field $\varphi$ acquires the effective mass squared $\Pi_\phi = 3\lambda\langle\varphi^2\rangle$. 
One may relate $\Pi_\phi = 3\lambda \langle\varphi^2\rangle$ to the number density of $\phi$-particles  in the following way:
\begin{eqnarray}\label{dispersion}
&&\Pi_\phi  \approx {3\lambda\over (2\pi)^3}\int {d^3 k\  n_k(\eta)\over \sqrt{k^2+ 3 \lambda \varphi^2(\eta) }}\nonumber\\ &\approx&  {3\lambda\over (2\pi)^3}\int {d^3 k\  n_k\over \sqrt{1.6 \tilde\varphi^2+ 3 \lambda \varphi^2 }} 
=     {3\lambda n_\phi(\eta) \over \sqrt{1.6\tilde\varphi^2+ 3 \lambda \varphi^2(\eta)} }\  .
\end{eqnarray}

Note, that this quantity is time-dependent. It oscillates;   its magnitude changes considerably  several times within a single oscillation of the inflaton field, and it also grows exponentially during the resonance. The number density of $\varphi$-particles  also oscillates and grows exponentially, but typically its oscillations are less wild than the oscillations of $\langle\varphi^2\rangle$. In the first approximation, we will neglect the oscillations of $n_\phi(\eta)$. Also, we are trying to find the time when the resonance terminates, and at that time   the average number density of particles $n_\phi$ becomes nearly constant.  It is still difficult to find an analytic solution for $\varphi_k$ with the time-dependent polarization operator (\ref{dispersion}), but one can easily find the solution  numerically.

The result of the combined investigation of the two effects discussed above shows that
the resonance on the leading modes $\varphi_k$ effectively terminates as soon as $\langle   \phi^2\rangle$ grows up to 
 \begin{equation}
  { \langle  \phi^2\rangle}    \approx 0.05\,  \tilde\phi^2 \ .
\label{end4}
\end{equation}
Note that even after this moment the resonance may continue for a while for the new modes which can be amplified in the restructured resonance band. However, this process is much less efficient.
 Thus,  in the pure $\lambda \phi^4$ theory the rapid development of the resonance ends
when  the dispersion of amplified fluctuations
 is about $20\%$ of the amplitude of the inflaton field,
which corresponds to only $0.2\%$ of the total energy.
This result is based on rather rough estimates neglecting rescattering. It is interesting, however, that it is in   complete agreement with the result  of the   lattice simulation
of the parametric resonance in the  theory
 $\lambda \phi^4$  \cite{KhTk961}.

 We should emphasize that there are several  specific reasons why the resonance in the particular case of the theory $\lambda\phi^4$ is relatively inefficient. First of all, the resonance band in this theory is very narrow and the characteristic exponent $\mu$ is extremely small. This is no longer the case when one considers, for example,  the theory describing a $\chi$-field with $g^2 = \lambda$ or with $g^2 = 2\lambda$. In these theories the characteristic exponent is much greater,   the resonance band is rather broad, and it begins at $k = 0$. As a result, it is much more difficult to shut down the resonance in such theories.

In the theories with a massive inflaton field there is an additional effect which makes the resonance more stable. Broad parametric resonance in such theories is stochastic, which makes it more difficult to shut down \cite{KLS97}. Now we are going to study what happens to the resonance in the conformally invariant theories if this invariance is broken by a small mass term. As we will see, stochastic resonance may appear in such theories as well.

\section{\label{massive}Preheating in the theory of a massive self-interacting  inflaton field}

In our previous paper \cite{KLS97} we investigated parametric resonance in the theory ${m^2\over 2}\phi^2 + {g^2\over 2} \phi^2\chi^2$. We have found that reheating can be efficient in this theory only if    $g\Phi \gg m$, where $\Phi$ is the amplitude of oscillations of the inflaton field. This amplitude is extremely large  immediately after inflation, $\phi \sim 10^{-1} M_p$, and later it decreases as 
\begin{equation}\label{Old1}
\Phi \sim {M_p\over 3mt} \ .
\end{equation}
Due to this decrease, the ratio   ${g\phi\over m}$ rapidly changes. As a result, the broad parametric resonance regime in this theory is a stochastic process, which we called {\it stochastic resonance}.

Here we studied the theory   ${\lambda\over 4}\phi^4 + {g^2\over 2} \phi^2\chi^2$
for various relations between the coupling constants $g^2$ and $\lambda$. In this theory the amplitude of the field $\phi$ also decreases in an expanding universe, but it does not make the resonance stochastic because all parameters of the resonance scale in the same way as $\Phi$ due to the conformal invariance. One may wonder, what is the relation between these two theories? Indeed, neither of these two theories is completely general. In the theory of the massive scalar field one may expect  terms $\sim  {\lambda\over 4}\phi^4$ to appear because of radiative corrections. On the other hand, in many realistic theories the effective potential is quadratic with respect to $\phi$ near the minimum of the effective potential.

To address this question, let us study the theory ${m^2\over 2}\phi^2 + {\lambda\over 4}\phi^4 + {g^2\over 2} \phi^2\chi^2$. One may expect that for  $\phi \gg { m\over\sqrt\lambda}$ parametric resonance in this theory occurs in the same way as in the model ${\lambda\over 4}\phi^4 + {g^2\over 2} \phi^2\chi^2$, whereas for  $\phi \ll  {m\over\sqrt\lambda}$ the resonance develops as in the theory   ${m^2\over 2}\phi^2 + {g^2\over 2} \phi^2\chi^2$. Let us check whether this is really the case, ignoring for simplicity the effects of backreaction of created particles, which is always possible in the beginning of the resonance regime. 

First of all, one should remember that at the beginning of the stage of oscillations in this theory one has $\Phi \sim 10^{-1} M_p$. Therefore there   are two basic possibilities. If ${ m\over\sqrt\lambda} \gg 10^{-1} M_p$, then the term ${\lambda\over 4}\phi^4 $ never plays any role in determining the frequency of oscillations of the field $\phi$. Also, in this regime the particles $\phi$ are not produced by parametric resonance, because the condition $\sqrt\lambda\phi > m$ (analogous to the condition $g\phi > m$ for the production of $\chi$-particles  \cite{KLS97}) is violated. In such a case $\chi$-particles can be produced if $10^{-1}g M_p \gg m$. The theory of this process is described in \cite{KLS97}; we do not have anything new to add here.

Another possibility, which we are going to study here in   more detail, is that ${ m\over\sqrt\lambda} \ll 10^{-1} M_p$. Then in the beginning the mass term ${m^2\over 2} \phi^2 $ does not affect the frequency of the oscillating scalar field $\phi$. Therefore, one could expect that as the amplitude $\Phi$ decreases from $10^{-1}M_p$ to ${m\over\sqrt\lambda}$, the theory of parametric resonance coincides with the one described in this paper.

However, for large $g^2\over \lambda$  the situation is more complicated. Even though the mass term for $10^{-1}M_p > \Phi \gg{m\over\sqrt\lambda}$ does not affect the frequency of oscillations, it may affect the nature of the broad parametric resonance by inducing an additional  rotation of the phase of the modes $\chi_k$.

The reason why the broad resonance in the theory ${m^2\over 2}\phi^2 + {g^2\over 2} \phi^2\chi^2$ was stochastic can be explained as follows. The $\chi$-particles are produced   when the field $\phi(t)$ comes close to the point $\phi =0$, which happens once during each time period $\Delta t ={ \pi\over m}$. During this time the phase of each mode $\chi_k$ grows approximately by ${g\Phi(t)\pi m^{-1}}$. During the next half of a period of an oscillation    it changes by  ${g\Phi(t+{ \pi\over m})\,\pi  m^{-1}}\approx  {g\Phi(t)\pi m^{-1}}+  {g\dot\Phi(t)\pi^2 m^{-2}}$. This destroys the phase coherence required for the ordinary resonance and makes the resonance stochastic   if $|{g\dot\Phi (t)\pi^2 m^{-2}}| \gtrsim 1$.

The condition for the stochastic resonance in the theory ${m^2\over 2}\phi^2 + {g^2\over 2} \phi^2\chi^2$ can be obtained from Eq. (\ref{Old1}):
\begin{equation}\label{}
\Phi   \gtrsim \sqrt{ mM_p\over    g} \ .
\end{equation}
In particular, for $\Phi = {m\over \sqrt\lambda}$ it gives ${g\over\sqrt\lambda } \gtrsim {\sqrt\lambda M_p\over   m}$. Note  that by our assumption ${\sqrt\lambda M_p\over   m} \gg 1$.

The generalization of this result for the theory ${m^2\over 2}\phi^2 + {\lambda\over 4}\phi^4 + {g^2\over 2} \phi^2\chi^2$ is straightforward, but the result is somewhat unexpected. As a rough estimate of the time $\Delta t$ one can take $\pi (2\lambda\Phi^2 + m^2)^{-1/2} = \pi (2\lambda\varphi^2 a^{-2}(t) + m^2)^{-1/2}$, where $\varphi\equiv \Phi a^{-1}(t)$ is the time-independent amplitude. The phase shift during this time is given by $g\varphi \pi (2\lambda\varphi^2  + m^2a^{2}(t))^{-1/2}$. Thus, for $m = 0$ this quantity is time-independent, and one can have a  regular stable resonance. In the limit $\Phi \gg {m\over\sqrt\lambda}$ one can represent the phase shift as ${g  \pi\over\sqrt{2\lambda} }(1  - {m^2a^{2}(t)\over 4\lambda\varphi^2})$. The change in this shift during one oscillation is $ {g  \pi^2 m^2 H \over 4\lambda^2 \Phi^3 } $, where $H = {\dot a\over a} = {\sqrt{2\pi \lambda} \Phi^2\over \sqrt 3 M_p}$.  This gives the following condition for stochastic resonance:
\begin{equation}\label{stoch}
\Phi \lesssim {g\over\sqrt\lambda}\, { \pi^2 m^2  \over 3\lambda  M_p}   . 
\end{equation} 
Again, for $\Phi = {m\over \sqrt\lambda}$ it gives ${g\over\sqrt\lambda } \gtrsim  {\sqrt\lambda M_p\over   m}$.

This conclusion is illustrated by Fig. 9, where we show the development of the resonance both for the massless theory with ${g^2\over \lambda} \sim 1700$, and for the theory with a small mass $m$. As we see, in the purely massless theory the logarithm of the number density $n_k$ for the leading growing mode increases linearly in time $x$, whereas in the presence of a   mass $m$, which we took to be much smaller than $\sqrt\lambda\phi$ during the whole process, the resonance becomes stochastic.

\begin{figure}[t]
\centering
\leavevmode\epsfysize=5.4cm \epsfbox{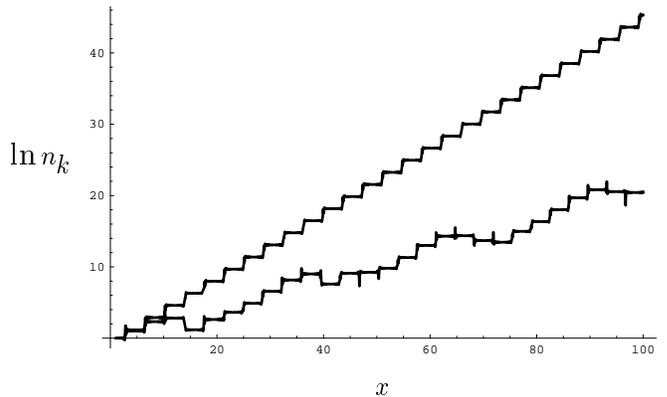}\\
\

\caption[Stoch]{\label{Stoch} Development of the resonance in the theory ${m^2\over 2}\phi^2 + {\lambda\over 4}\phi^4 + {g^2\over 2} \phi^2\chi^2$  for ${g^2\over \lambda} = 5200$. The upper curve corresponds to the massless theory, the lower curve describes stochastic resonance with a theory with a  mass $m$ which is chosen to be much smaller than  $\sqrt\lambda  \phi $ during the whole period of calculations. Nevertheless, the presence of a small mass term completely changes the development of the resonance.}
\end{figure}

In fact, the development of the resonance is rather complicated even for smaller  ${g^2\over \lambda}$. The resonance for a massive field with $m \ll \sqrt\lambda\phi$ in this case is not   stochastic, but it may consist  of stages of regular resonance separated by the stages without any resonance, see Fig. 10. 

\begin{figure}[t]
\centering
\leavevmode\epsfysize=5.5cm \epsfbox{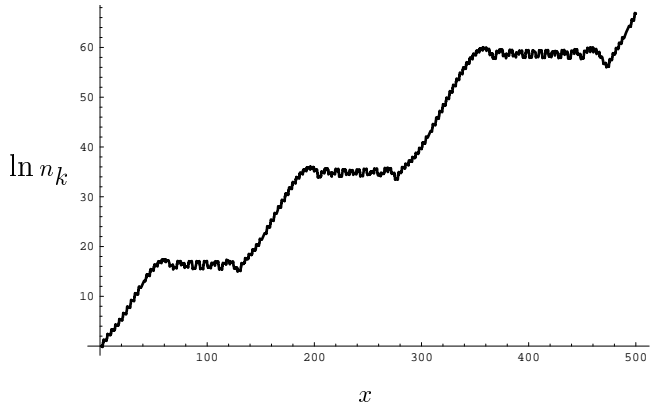}\\
\

\caption[ResMass]{\label{ResMass} Development of the resonance in the theory ${m^2\over 2}\phi^2 + {\lambda\over 4}\phi^4 + {g^2\over 2} \phi^2\chi^2$ with  $ m^2  \ll  \lambda  \phi^2$ for ${g^2\over \lambda} = 240$. In this particular case the resonance is not stochastic.  As time $x$ grows, the relative   contribution of the mass term to the equation describing the resonance also grows. This shifts the mode from one instability band to another.}  
\end{figure}

Thus we see that the presence of the mass term ${m^2\over 2}\phi^2$ can modify the nature of the resonance even if this term is much smaller than ${\lambda\over 4}\phi^4 $. This is a rather unexpected conclusion, which is an additional manifestation of the nonperturbative nature of preheating. This subject   deserves separate investigation

\begin{figure}[t]
\centering
\leavevmode\epsfysize=5.9cm \epsfbox{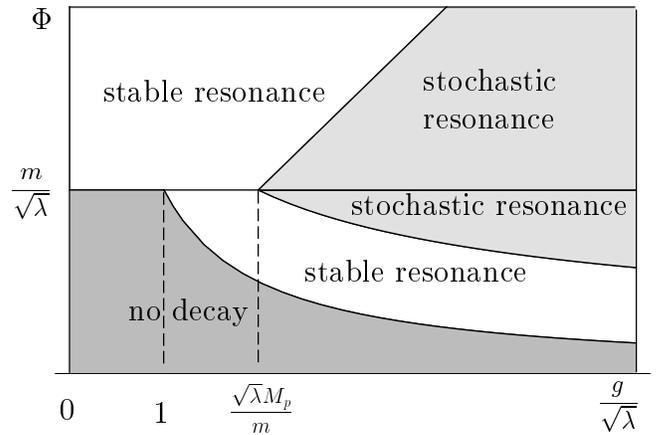}\\
\

\caption[ResRanges]{\label{ResRanges} Schematic representation of different regimes which are possible in the theory ${m^2\over 2}\phi^2 + {\lambda\over 4}\phi^4 + {g^2\over 2} \phi^2\chi^2$ for ${ m\over\sqrt\lambda} \ll 10^{-1} M_p$ and for various relations between $g^2$ and $\lambda$ in an expanding universe. The theory developed in this paper describes the resonance in the white area above the line $\Phi = {m\over \sqrt\lambda}$. The theory of preheating for $\Phi < {m\over \sqrt\lambda}$ is given in \cite{KLS97}. A complete decay of the inflaton is possible only if additional interactions are present in the theory which allow  one inflaton particle to decay to several other particles, for example, an interaction with fermions $   \bar\psi \psi \phi$.}
\end{figure}

Different regimes of parametric resonance in the theory ${m^2\over 2}\phi^2 + {\lambda\over 4}\phi^4 + {g^2\over 2} \phi^2\chi^2$ are shown in Fig. \ref{ResRanges}. We suppose that immediately after inflation the amplitude $\Phi$ of the oscillating inflaton field is greater than ${m\over\sqrt\lambda}$. If ${g\over \sqrt\lambda} \lesssim {\sqrt\lambda M_p\over   m}$, the $\chi$-particles are produced in the regular stable resonance regime until the amplitude $\Phi(t)$ decreases to ${m\over \sqrt\lambda}$, after which the resonance occurs as in the theory ${m^2\over 2}\phi^2   + {g^2\over 2} \phi^2\chi^2$ \cite{KLS97}. The resonance never becomes stochastic.

If  ${g\over \sqrt\lambda} \gtrsim {\sqrt\lambda M_p\over   m}$, the resonance originally develops as in the conformally invariant theory ${\lambda\over 4}\phi^4 + {g^2\over 2} \phi^2\chi^2$, but with a decrease of $\Phi(t)$ the resonance becomes stochastic. Again, for 
$\Phi(t) \lesssim {m\over \sqrt\lambda}$  the resonance occurs as in the theory ${m^2\over 2}\phi^2   + {g^2\over 2} \phi^2\chi^2$. In all cases the resonance eventually disappears when the field $\Phi(t)$ becomes sufficiently small. As we already mentioned in \cite{KLS,KLS97},  reheating in this class of models can be complete only if there is a symmetry breaking in the theory, i.e. $m^2 < 0$, or if one adds interaction of the field $\phi$ with fermions. In both cases the last stages of reheating are described by perturbation theory \cite{DL,PERT}.

\begin{figure}[t]
\centering
\leavevmode\epsfysize=5.9cm \epsfbox{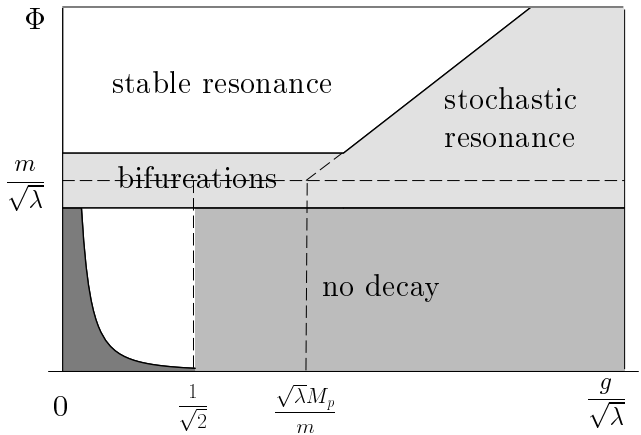}\\
\

\caption[RRS]{\label{RRS} Schematic representation of different regimes which are possible in the theory $-{m^2\over 2}\phi^2 + {\lambda\over 4}\phi^4 + {g^2\over 2} \phi^2\chi^2$. White regions correspond to the regime of a regular stable resonance, a small dark region in the left corner near the origin corresponds to the perturbative decay $\phi \to \chi\chi$. Unless additional interactions are included (see the previous Figure), a complete decay of the inflaton field is possible only in this small area.}
\end{figure}

Adding fermions does not alter the description of the stage of parametric resonance. Meanwhile the change of sign of $m^2$ does lead to substantial changes in the theory of preheating,   see Fig. \ref{RRS}.  We will investigate preheating in the theory $-{m^2\over 2}\phi^2 + {\lambda\over 4}\phi^4 + {g^2\over 2} \phi^2\chi^2$ in a separate publication \cite{PhaseTr}. Here we will briefly describe the structure of the resonance for various $g^2$ and $\lambda$ neglecting effects of backreaction. This will give us a more general perspective on the theory of reheating.

First of all,  at $\Phi \gg {m/\sqrt \lambda}$ the  field $\phi$ oscillates in the same way as in the massless theory ${\lambda\over 4}\phi^4 + {g^2\over 2} \phi^2\chi^2$. Moreover, the condition for the resonance to be stochastic remains the same as before: $\Phi \lesssim {g\over\sqrt\lambda}\, { \pi^2 m^2  \over 3\lambda  M_p}$, see Eq. (\ref{stoch}).

However, as soon as the amplitude $\Phi$ drops down to ${m\over \sqrt\lambda}$, the situation changes dramatically. First of all, depending on the values of parameters the field rolls to one of the minima of its effective potential  at $\phi = \pm {m\over \sqrt \lambda}$. The description of this process is rather complicated. Depending on the values of parameters and on the relation between $\sqrt{\langle\phi^2\rangle}$, $\sqrt{\langle\chi^2\rangle}$ and $\sigma  \equiv {m \over \sqrt\lambda}$, the universe may become  divided   into domains with $\phi = \pm \sigma$, or it may end up in a single state with a definite sign of $\phi$. We will describe this bifurcation  period in \cite{PhaseTr}. After this transitional period the field $\phi$ oscillates near the   minimum of the effective potential at $\phi = \pm {m\over \sqrt \lambda}$ with an amplitude $\Phi \ll \sigma = {m\over \sqrt\lambda}$.   These oscillations lead to parametric resonance with $\chi$-particle production which can be (approximately) described as a narrow resonance in the first instability band of the Mathieu equation with $A_k = 4{k^2+g^2\sigma^2 \over m^2 }$, $q = {4g^2\sigma\Phi \over
 m^2} $. For definiteness we will consider here the regime $\lambda^{3/2}M_p < m \ll \lambda^{1/2}M_p$. The resonance in this instability band is possible only if ${g^2\over  \lambda} < {1\over 2}$; the resonance in higher instability bands is very inefficient and rapidly shuts down due to the expansion of the universe.  Using the results of \cite{KLS97} one can show that the resonance in the first band also terminates at $\Phi < {\lambda m^2\over g^4 M_p}$. By taking the upper limit of this inequality at $\Phi \sim {m\over\sqrt\lambda}$ one concludes that this resonance is possible only for ${g\over\sqrt\lambda} \gtrsim  \bigl({m\over\sqrt\lambda M_p}\bigr)^{1/4}$.  (The resonance may terminate somewhat earlier if the particles produced by the parametric resonance give a considerable contribution to the energy density of the universe.) However, this is not the end of reheating, because the perturbative decay of the inflaton field remains possible. It occurs with the decay rate $ \Gamma ( \phi \to \chi \chi) = { g^4 m\over 8
\pi \lambda}$. This is the process which is responsible for the last stages of the decay of the inflaton field. It occurs only if one $\phi$-particle can decay into two $\chi$-particles, which implies that $ {g^2\over \lambda}  < {1\over 2}$.  

\section{\label{disc} Discussion }

In this paper we investigated the development of parametric resonance in the conformally invariant theories of the type of ${\lambda\over 4}\phi^4 + {g^2\over 2} \phi^2\chi^2$. We have found that the development of parametric resonance in these theories does not depend on the expansion of the universe, and can be classified in terms of the ratio $g^2/\lambda$. This ratio determines the structure of the stability/instability bands for the equations describing the resonance.

We have found that the behavior of the resonance with respect to   $\chi$-particle production is a non-monotonic function of $g^2/\lambda$. For example, for $g^2 = \lambda$ and for $g^2 = 3\lambda$ equation for the perturbations of the field $\chi$ has only one instability band, for ${g^2 \over \lambda} ={n(n+1) \over 2}$ there is only a finite number of instability bands, whereas for all other values of  ${g^2 \over \lambda}$ the number of instability bands is infinite.  

It is interesting that  $\chi$-particle production is   least efficient for $g^2 \ll \lambda$ and for $g^2 = 3\lambda$. For example, the characteristic exponent $\mu_{\rm max}$ for $g^2 = 2\lambda$ and for   $g^2 = 8\lambda$ is almost 7 times greater than $\mu_{\rm max}$ for $g^2 = 3\lambda$, see Fig. 6. Meanwhile the characteristic exponent for the production of $\phi$-particles in the theory ${\lambda\over 4}\phi^4$ coincides with that of the field $\chi$ for $g^2 = 3\lambda$. Therefore     $\chi$-particle production is typically more efficient than the production of $\phi$-particles (unless $g^2 \ll \lambda$). The non-monotonic dependence of $\mu$ on the ratio ${g^2\over \lambda}$ suggests that there exists an ``unnatural selection'' rule: The particles which are especially intensively produced during preheating are not the ones which have the strongest coupling to the inflaton field, but those for which the characteristic exponent $\mu$ is the greatest.

In the conformally invariant theories the expansion of the universe does not hamper the resonance, so it ends only due to the backreaction of the produced particles. There are several different mechanisms which may terminate the parametric resonance. First of all, creation of particles leads to a decrease in the amplitude of   oscillations of the field $\varphi = a\phi$, which otherwise would remain constant. This leads to a proportional decrease in the frequency of oscillations in terms of the conformal time $\eta$, which may shift the position of the  instability band towards smaller momenta. There is also an opposing effect which increases   the frequency of oscillations due to the interaction of the homogeneous inflaton field with the produced particles. Finally, quantum fluctuations of the fields $\phi$ and $\chi$ acquire  contributions to their masses, which changes their spectra. A combination of all these effects leads to restructuring of the instability bands.   This terminates the amplification  of the  leading modes   which have been growing from the very beginning of preheating. Additionally, one may envisage effects related to rescattering of produced particles, which may terminate the resonance even somewhat earlier. In this respect it is interesting that our estimates ignoring the process of rescattering give   results which are in a very good numerical agreement with the results of computer simulations of reheating in the theory $\lambda\phi^4$ performed in \cite{KhTk961} where all of these effects including rescattering have been taken into account. 

Rescattering may be more important for $g^2 \gg \lambda$ \cite{rt,Prokopec}.
However, in this regime one may need to take into account possible small mass terms which should be present in realistic versions of the theory. As we have found, for $g^2 \gg \lambda$ these mass terms lead to a radical change in the structure of the resonance not at $\Phi \lesssim m/\sqrt\lambda$, as one could naively expect, but much earlier, at $\Phi \lesssim {g\over\sqrt\lambda}\, { \pi^2 m^2  \over 3\lambda  M_p}$. In this regime the resonance becomes stochastic, the effective width of the resonance band increases,   making it much more stable with respect to various backreaction effects including rescattering \cite{KLS97}.

We should emphasize again  that preheating is but the first stage of reheating, which does not lead  to a complete decay of the inflaton field in any models which we studied so far. The last stages of preheating are always described by the perturbation theory \cite{DL}, which will be developed further in our subsequent publication \cite{PERT}. To illustrate this point, we  described the development of the parametric resonance in the general class of models with the effective potential $V(\phi,\chi) = \pm {m^2\over 2}\phi^2 + {\lambda\over 4}\phi^4 + {g^2\over 2} \phi^2\chi^2$. We have found that in these theories (without any other fields being added) the inflaton field can  completely decay only if the sign   of the   term ${m^2\over 2}\phi^2$ is negative, which corresponds to spontaneous symmetry breaking. Moreover, this process completes only    for ${g^2\over \lambda}  < {1\over 2}$, see Figs. \ref{ResRanges} and \ref{RRS}.  

A complete inflaton decay is possible for ${g^2\over \lambda}  > {1\over 2}$ as well, even without   spontaneous symmetry breaking, but only if the inflaton field has some other interactions, such as an interaction with   fermions $\bar\psi\psi\phi$ with mass $m_\psi <m/\sqrt 2$ \cite{PERT}. This conclusion  implies that the decay of the inflaton field  is by no means automatic even if it is heavy and strongly interacts with other fields.   Generically,  the inflaton field accumulates an enormously large energy density, which can be completely released only if it interacts with other particles in a very specific way \cite{KLS}.

To understand how these results may change our point of view on the thermal history of the universe, let us suppose for a moment that the inflaton field does not have any   interactions   with light fermions, and  that it has an effective potential $-{m^2\over 2}\phi^2 + {\lambda\over 4}\phi^4+ {g^2\over 2} \phi^2\chi^2$ with $\lambda \sim g^2 \sim 10^{-13}$ and with a small mass $m \sim 10^2$ GeV protected by supersymmetry. Then the final stage of reheating of the universe will begin only after the symmetry breaking in this theory, and   the reheating temperature estimated in accordance with \cite{KLS97} will be smaller than $10^2-10^3$ GeV. In such a theory the electroweak phase transition may never happen, or it may occur in an entirely different way. From the end of inflation until the symmetry breaking and the final stage of reheating, the universe will remain  far away from thermal equilibrium, and various nonthermal phase transitions and explosive processes of particle production may occur. In such a model one should reconsider all issues related to the primordial gravitino problem, moduli field problem, baryogenesis, etc.

The main conclusion of our investigation can be formulated as follows. The first stages of the process of the inflaton decay may occur much more efficiently than was previously thought, due to the effect of parametric resonance. The last stage  of this process may be completely inefficient even if the coupling of the inflaton field to matter is very strong, or it may be efficient only in a very narrow range of parameters, see Figs. \ref{ResRanges} and \ref{RRS}. As a result, the complete thermal (and nonthermal) history of the universe   in the context of the inflationary universe scenario may be dramatically different from the standard lore of the hot Big Bang cosmology.

\bigskip
\section*{Acknowledgments}
The authors are grateful to Igor Tkachev   for useful discussions.
 This work was supported  by NSF grant AST95-29-225.
 The work by A.L. was also supported   by NSF
grant PHY-9219345.
A.S. was supported   by the Russian
Foundation
for Basic Research, grant 96-02-17591.
A.L and A.S. thank the Institute for Astronomy, University of
Hawaii for hospitality.

\section*{Appendix}

Here we   show how one can derive Eq. (\ref{mu2}) or (\ref{mu3}) for the characteristic exponent $\mu_k$ from the analytic solution (\ref{form1}).
For simplicity, we will consider here the case   $g^2 = \lambda$. Eq. (\ref{form1}) describes both solutions, $X_1(z)$ and $X_2(z)$. The resonant solution $X(z)$ consists of four monotonic parts within 
a single period of the inflaton oscillation, see Fig.  \ref{fig2a}. It turns out that at different quarters of the period either  $X_1(z)$ or $X_2(z)$ correspond to the exponentially growing solution.   Indeed, the square of the resonant solution within the first quarter of a period is
\begin{equation}\label{A1}
X_{1/4}^2(z) = X^2_0\exp\biggl[\int\limits_0^z{dz\over M_1(z)}\Bigl(1-{C\over\sqrt{z(1-z^2)}}\Bigr)\biggr] ,
\end{equation}
where $M_1(z)$ is given by (\ref{m1}),   $C$ is given by (\ref{C1}), and $X_0^2$ is the square of the resonant solution in the beginning of the period when $z = 0$.

Within the second quarter  one has
\begin{equation}\label{A2}
X_{1/2}^2(z) = X_{1/4}^2(1)\, \exp\left[\int\limits_1^z{dz\over M_1(z)}\Bigl(1+{C\over\sqrt{z(1-z^2)}}\Bigr)\right] .
\end{equation}

 Then the value  of $X^2$ after   half of a period is  
\begin{equation}\label{A3}
 X_{1/2}^2(z= 0) 
 =  X_0^2(z = 1)  \exp\Bigl(-2C \int\limits_0^1{dz\over M_1(z)\sqrt{z(1-z^2)}} \Bigr) ,
\end{equation}
where the integral is understood as its   principal value.
The resonant solution has the generic form $X(z(x)) = P(z(x)) e^{\mu x}$, where $P(z)$ is a periodic function. Since $P$ has a period equal to   half of the period of the inflaton oscillation, Eq. (\ref{A3}) is sufficient to find $\mu$:
\begin{equation}\label{A4}
{\mu T\over 2} = -C \int\limits_0^1{dz\over M_1(z)\sqrt{z(1-z^2)}} > 0 .
\end{equation}
The integral in this equation can be reduced to $I(\kappa^2)$ given by (\ref{I}):
\begin{equation}\label{A5}
  -  \int\limits_0^1{dz\over M_1(z)\sqrt{z(1-z^2)}}  =   \int\limits_0^{r/2}{d\theta\sin^{1/2}\theta\over 1+2\kappa^2 \sin\theta}  \equiv I(\kappa^2).
\end{equation}


\begin{thebibliography}{999}
\bibitem{KLS}
L. A. Kofman, A. D. Linde,  and A. A. Starobinsky,
	Phys. Rev. Lett. {\bf 73}, 3195 (1994).

\bibitem{KLS97} L. A. Kofman, A. D. Linde,  and A. A. Starobinsky,
``Towards the Theory of Reheating After Inflation,'' preprint IfA-97-28, SU-ITP-97-18 (1997), hep-ph/9704452.

\bibitem{Shtanov} Y. Shtanov, J. Traschen, and R. Brandenberger,
Phys.Rev.  D {\bf 51}, 5438 (1995).


\bibitem{Kof96} L. Kofman, In:
{\it Relativistic Astrophysics: A Conference in Honor of Igor
 Novikov's 60th Birthday.}
Copenhagen 1996. Eds. B. Jones and D. Marcovic, Cambridge University Press,
astro-ph/9605155.

\bibitem{KhTk961} S.Yu. Khlebnikov and I.I. Tkachev,
Phys. Rev. Lett. {\bf 77}, 219 (1996), hep-ph/9603378;  

\bibitem{Boyan95} D. Boyanovsky, H.J. de Vega, R. Holman,
D.S. Lee, and A. Singh, Phys. Rev. D {\bf 51}, 4419 (1995).

\bibitem{Boyan96} D. Boyanovsky, H.J. de Vega, R. Holman,
D.S. Lee, and A. Singh,  and J. F. J. Salgado, Phys. Rev. D {\bf 54}, 7570
(1996).

\bibitem{Kaiser97} D.I. Kaiser, HUTP-97-A005, hep-ph/9702244.

\bibitem{rt} A. Riotto and I. Tkachev, Phys. Lett. {\bf B385}, 57
(1996), hep-ph/9604444.


\bibitem{Prokopec} T. Prokopec and T. G. Roos, Phys. Rev. D {\bf 55}, 3768
(1997), hep-ph/9610400.

\bibitem{Turner}   M.S. Turner, Phys. Rev. {\bf D28}, 1243 (1983).

 
\bibitem{erd}    A. Erdelyi,  ed.
 {\it Higher Transcendental Functions}, (Bateman Manuscript Project)
vol.3 (NY: McGraw-Hill, 1955).


\bibitem{AS} M. Abramowitz and I. Stegun, {\it Handbook of
Mathematical Functions}, p.685. (Dover, New York,1965).


\bibitem{KLS96} L. A. Kofman, A. D. Linde,  and A. A. Starobinsky,
	Phys. Rev. Lett. {\bf 76}, 1011 (1996); I. Tkachev, Phys. Lett. {\bf B376},  35 (1996).


\bibitem{CommSht}One could try  to investigate the resonance in the theory  ${1 \over 4 } \lambda \phi^4$
using an iteration series in   the 
small parameter $\lambda$ \cite{Shtanov}.  Unfortunately, this approach
is  not very informative since the actual parameter  defining the strength  of the resonance
is not $\lambda$  but $g^2/\lambda=3$, which is not small.



\bibitem{DL} A.D. Dolgov and A.D. Linde, Phys. Lett. {\bf 116B}, 329
(1982); L.F. Abbott, E. Fahri and M. Wise, Phys. Lett. {\bf
117B}, 29 (1982); A.A. Starobinsky, in: {\it Quantum Gravity, Proc.
of the Second Seminar ``Quantum theory of Gravity'' (Moscow, 13-15
Oct. 1981)}, eds. M.A. Markov and P.C. West (Plenum, New York,
1984), p. 103.

\bibitem{PERT} L. A. Kofman, A. D. Linde, and A. A. Starobinsky, ``Reheating
After Inflation: Perturbation Theory,'' in preparation.

\bibitem{PhaseTr} S. Khlebnikov, L. A. Kofman, A. D. Linde, A. A. Starobinsky,
and I. Tkachev, in preparation.


 \end{thebibliography}
\end{document}